\documentclass[10pt]{article}
\usepackage{epsfig}
\usepackage{amsmath}
\usepackage{cite}
\def\CA{C_A}

\def\NF{N_F}
\def\mom#1{\langle #1 \rangle}

\def\d{\hbox{d}}

\def\ln{\hbox{ln}}

\begin{document}
\unitlength1cm
\begin{titlepage}
\vspace*{-1cm}
\begin{flushright}
ZU-TH 14/09
\end{flushright}
\vskip 3.5cm

\begin{center}
{\Large\bf Hadronization effects in event shape moments}
\vskip 1.cm
{\large  T.~Gehrmann, M.~Jaquier and G.~Luisoni}
\vskip .7cm
{\it
Institut f\"ur Theoretische Physik, Universit\"at Z\"urich,
Winterthurerstrasse 190,\\ CH-8057 Z\"urich, Switzerland}
\end{center}
\vskip 2cm

\begin{abstract}
We study the moments of hadronic event shapes in $e^+e^-$ annihilation within
the context of  next-to-next-to-leading order
(NNLO) perturbative QCD predictions combined with non-perturbative power
corrections in the dispersive model. This model is extended to match upon the
NNLO perturbative prediction.
The resulting theoretical expression  has been compared to experimental data from JADE and OPAL, and a new value for $\alpha_s(M_Z)$ has been determined, as well as of the average coupling  $\alpha_0$
in the non-perturbative region below $\mu_I=2$ GeV within the
dispersive model:
\begin{align}
\alpha_s(M_Z)&=0.1153\pm0.0017(\mathrm{exp})\pm0.0023(\mathrm{th}),\nonumber \\
\alpha_0&=0.5132\pm0.0115(\mathrm{exp})\pm0.0381(\mathrm{th})\,,\nonumber 
\end{align}
The precision of the $\alpha_s(M_Z)$ value has been improved in comparison to
the previously available next-to-leading order analysis. 
We observe that the resulting power corrections are considerably larger 
than those estimated from hadronization models in multi-purpose event 
generator programs.
\end{abstract}
\end{titlepage}
\newpage

\section{Introduction}

Event shape variables measure geometrical properties of hadronic final states
at high energy particle collisions. They have been studied extensively
at $e^+e^-$ collider experiments, which provided a wealth of data
at a variety of centre-of-mass energies. Exploiting this large energy
range, one can attempt to disentangle perturbative and non-perturbative
contributions  (which scale differently with increasing
energy) to event shape observables.

Apart from distributions of these observables,
one can also study mean values and higher moments.
The $n$th moment of an event shape observable $y$ is
defined by
\begin{equation}
\mom{y^n}=\frac{1}{\sigma_{\rm{had}}}\,\int_0^{y_{\rm{max}}} y^n
 \frac{\d\sigma}{\d y} \d y \;,
\end{equation}
where $y_{\mathrm{max}}$ is the kinematically allowed upper limit of the
observable.
Moments were measured for a variety of different event shape variables in the
past. The most common observables $y$ of three-jet type are:
thrust $T$~\cite{farhi} (where
moments of $y=(1-T)$ are taken), the heavy jet mass $\rho = M_H^2/s$~\cite{mh},
the $C$-parameter~\cite{c}, the wide and total jet broadenings $B_W$ and
$B_T$~\cite{bwbt}, and the
three-to-two-jet transition parameter in the Durham algorithm
$Y_3$~\cite{durham}.
Definitions for all observables are given in, for example,
Ref.~\cite{GehrmannDeRidder:2007hr}.
Moments with $n\geq 1$ have been measured by several experiments, most extensively by
{\small JADE}~\cite{MovillaFernandez:1997fr,jadenew}
and {\small OPAL}~\cite{Abbiendi:2004qz},
but also by {\small DELPHI}~\cite{Abreu:1999rc} and L3~\cite{Achard:2004sv}.
A combined analysis of JADE and OPAL results has been performed in Ref.~\cite{Kluth:2000km}.

As the calculation of moments involves an integration over the full phase
space,  they offer a way of comparing to data which is complementary to the use
of distributions, where in general cuts on certain kinematic regions are
applied. Furthermore, the two extreme kinematic limits -- two-jet-like events
and multi-jet-like events -- enter with different weights in each moment:  the
higher the order $n$ of the moment, the more it becomes sensitive to the
multi-jet region. Therefore it is particularly interesting to study the NNLO
corrections to higher moments of event shapes, as these corrections  should
offer a better description of the multi-jet region due to the inclusion of
additional radiation at parton level.

Moments are particularly attractive in view of studying
non-perturbative hadronization
corrections to event shapes. In event shape distributions, one typically
corrects for hadronization effects by using generic Monte Carlo
event simulation programs. A recent study, carried out in the context
of a precision determination of the strong coupling constant from
event shape distributions~\cite{newas}, revealed large discrepancies between
the standard event simulation programs used at LEP~\cite{Sjostrand:2000wi,herwigpythia}
on one hand and
more modern generators~\cite{hw++etal},
which incorporate recent theoretical advances, on the other hand. In the
event shape distributions, it is very difficult to disentangle
hadronization corrections empirically, since they typically result in a
distortion of the distribution, which can not be unfolded in a straightforward
manner.

In event shape moments, one expects the hadronization
corrections to be additive, such that
they can be divided into a perturbative
and a non-perturbative contribution,
\begin{equation}
\mom{y^n}=\mom{y^n}_{\rm{pt}}+\mom{y^n}_{\rm{np}}\;,
\label{mom}
\end{equation}
where the non-perturbative contribution accounts for hadronization effects.
Based upon the calculation of
next-to-next-to-leading order (NNLO) QCD corrections to the
event shape distributions, which became available
recently~\cite{GehrmannDeRidder:2007hr,GehrmannDeRidder:2007bj,GehrmannDeRidder:2007jk,Weinzierl:2009ms,Weinzierl:2009nz,weinzierlnew},
the perturbative contribution
to event shape moments is now known
to NNLO~\cite{ourmom,weinzierlmom}.  The
non-perturbative part is suppressed by powers of $\lambda_p/Q^{p}\; (p\geq 1)$,
where  $Q\equiv \sqrt{s}$ is the centre of mass energy  and $\lambda_1$ is of
the order of  $\Lambda_{QCD}$.  The functional form of $\lambda_p$ has been
discussed quite extensively in the literature,  but as this parameter is closely
linked to  non-perturbative effects, it cannot be fully derived from first
principles.

In this work, we use the dispersive model derived in
Ref.~\cite{Dokshitzer:1995qm,Dokshitzer:1997ew,Dokshitzer:1998pt,Dokshitzer:1998qp}
to compute hadronization corrections to event shape moments. This model
provides analytical predictions for the power corrections, and introduces
only a single new parameter $\alpha_0$, which can be interpreted
as the average strong coupling in the non-perturbative region. This model
has been used extensively in combination with NLO QCD perturbative
calculations to study event shape
 moments~\cite{Abbiendi:2004qz,jade,pahl,others}.
To combine the dispersive model with the perturbative prediction at NNLO QCD,
we  extended its analytical expressions to compensate for all scale-dependent
terms at this order. By comparing the newly derived expressions with
experimental data on event shape moments, we perform a combined determination
of the perturbative strong coupling constant $\alpha_s$ and the
non-perturbative parameter $\alpha_0$. Compared to previous results at NLO,
we observe that inclusion of NNLO effects results in
a considerably improved consistency in the parameters determined from
different shape variables, and in a substantial reduction of the error on
$\alpha_s$.

In Section~\ref{sec:power}, we outline the structure of perturbative
and non-perturbative contributions to event shape moments. The predictions
of the dispersive model to power corrections are extended to NNLO in
Section~\ref{sec:disp}, and used to extract $\alpha_s$ and $\alpha_0$
from experimental data in Section~\ref{sec:exp}. In Section~\ref{sec:MCcomparison} the results obtained within the dispersive model are compared to those from multi-purpose event generator programs.

\section{Power corrections to event shape moments}
\label{sec:power}
Non-perturbative power corrections can be related to infrared renormalons
in the perturbative QCD expansion for the event shape
variable~\cite{Manohar:1994kq,Webber:1994cp,Korchemsky:1994is,Dokshitzer:1995zt,Akhoury:1995sp,Dokshitzer:1995qm,Nason:1995hd,Dokshitzer:1997ew}.
The analysis of infrared renormalon ambiguities suggests power corrections of the
form $\lambda_p/Q^{p}$, but cannot make unique predictions for $\lambda_p$:
it is only the sum of perturbative and non-perturbative contributions in
(\ref{mom}) that becomes well-defined~\cite{Beneke:1998ui}.
Different ways to regularise the IR renormalon singularities
have been worked out in the literature \cite{Dokshitzer:1997iz,Korchemsky:2000kp,Belitsky:2001ij,Gardi:1999dq,Campbell:1998qw,Dasgupta:2003iq}.

One approach is to introduce an  IR cutoff $\mu_{I}$ and to replace the strong
coupling constant below the scale $\mu_{I}$ by an effective coupling
such that the integral of the coupling below $\mu_{I}$ has a finite value\cite{Dokshitzer:1995qm,Dokshitzer:1997ew,Dokshitzer:1998pt,Dokshitzer:1998qp}
\begin{equation}
\frac{1}{\mu_I}\int_0^{\mu_I} dQ \,\alpha_{\rm{eff}}(Q^2)=\alpha_0(\mu_I)\;.
\label{alpha0}
\end{equation}
This dispersive model for the strong coupling leads to a shift in the distributions
\begin{equation}
\frac{\d\sigma}{\d y}(y)=\frac{\d\sigma_{\rm{pt}}}{\d y}\,(y-a_y\,P)\;,
\label{eq:disp}
\end{equation}
where the numerical factor $a_y$ depends on the event shape
and is listed in Table~\ref{a_y_coeff}, while
${P}$ is believed to be universal (universality breaking terms
arise from hadron mass effects~\cite{Salam:2001bd}
 in the moments of $\rho$, an estimate on these effects can be obtained 
from general-purpose event generator programs, 
e.g.\ from PYTHIA~\cite{Sjostrand:2000wi})
 and scales with the CMS energy like  $\mu_I/Q$.
\begin{table}[b]
\begin{center}
\begin{tabular}{lllllll}
\hline
event shape observable & $1-T$ & $C$ & $Y_3$ & $\rho$ & $B_T$ & $B_W$\\ \hline
$a_y$ & 2 & 3$\pi$ & 0 & 1 & 1& $\frac{1}{2}$\\ \hline
\end{tabular}
\end{center}
\caption{The $a_y$ coefficients of the non-perturbative event shape moment prediction}\label{a_y_coeff}
\end{table}

By inserting (\ref{eq:disp}) into the definition of the moments, one obtains:
\begin{align} \langle y^n \rangle &= \int^{y_\mathrm{max}}_0\mathrm{d}yy^n\frac{1}{\sigma _\mathrm{had}}\frac{\mathrm{d}\sigma}{\mathrm{d}y}(y) \\
&=\int^{y_\mathrm{max}-a_yP}_{-a_yP}\mathrm{d}y(y+a_yP)^n\frac{1}{\sigma _\mathrm{had}}\frac{\mathrm{d}\sigma_\mathrm{pt}}{\mathrm{d}y}(y) \\
&\approx \int^{y_\mathrm{max}}_0\mathrm{d}y(y+a_yP)^n\frac{1}{\sigma _\mathrm{had}}\frac{\mathrm{d}\sigma_\mathrm{pt}}{\mathrm{d}y}(y)\label{eq:pc}
\end{align}
discarding the integration over the kinematically forbidden values of $y$. This
leads to the the non-perturbative predictions for the moments of $y$:
\begin{align} 
\langle y^1 \rangle &= \langle y^1 \rangle_\mathrm{pt} + a_yP,\nonumber \\
\langle y^2 \rangle &= \langle y^2 \rangle_\mathrm{pt} +2\langle y^1 \rangle_\mathrm{pt} (a_yP)+(a_yP)^2, \nonumber \\
\langle y^3 \rangle &= \langle y^3 \rangle_\mathrm{pt} +3\langle y^2 \rangle_\mathrm{pt} (a_yP)+3\langle y^1 \rangle_\mathrm{pt} (a_yP)^2 +(a_yP)^3,
\nonumber \\
\langle y^4 \rangle &= \langle y^4 \rangle_\mathrm{pt} +4\langle y^3 \rangle_\mathrm{pt} (a_yP)+6\langle y^2 \rangle_\mathrm{pt} (a_yP)^2 + 4\langle  y^1\rangle_\mathrm{pt}(a_yP)^3 +(a_yP)^4, \nonumber \\
\langle y^5 \rangle &= \langle y^5 \rangle_\mathrm{pt} +5\langle y^4 \rangle_\mathrm{pt} (a_yP)+10\langle y^3 \rangle_\mathrm{pt} (a_yP)^2 + 10\langle  y^2 \rangle_\mathrm{pt}(a_yP)^3 +5\langle y^1 \rangle_\mathrm{pt}(a_yP)^4+(a_yP)^5 \label{eq9}
\end{align}

It should be noted that the multiplicative power correction in~(\ref{eq:pc}) is considered to be accurate to $1/Q$. For $n\geq 2$, the evaluation~\eqref{eq9} yields also higher powers of $P$, which are formally of higher order in inverse powers of $Q$. Contributions with the same scaling behaviour could equally come from subleading power corrections in $P$. Compared to the terms above, these subleading power corrections would be weighted with higher perturbative moments, and are thus suppressed numerically.

The  perturbative contribution to $\mom{y^n}$ is given up to NNLO in terms of the dimensionless coefficients $\bar{{\cal A}}_{y,n}$, $\bar{{\cal B}}_{y,n}$ and $\bar{{\cal C}}_{y,n}$ as:
\begin{eqnarray}
\mom{y^n}_\mathrm{pt} (s,\mu^2) &=&
\left(\frac{\alpha_s(\mu)}{2\pi}\right) \bar{{\cal A}}_{y,n} +
\left(\frac{\alpha_s(\mu)}{2\pi}\right)^2 \left(
\bar{{\cal B}}_{y,n}+ \bar{{\cal A}}_{y,n} \beta_0
\log\frac{\mu^2}{s} \right)
\nonumber \\ &&
+ \left(\frac{\alpha_s(\mu)}{2\pi}\right)^3
\bigg(\bar{{\cal C}}_{y,n}+ 2 \bar{{\cal B}}_{y,n}
 \beta_0\log\frac{\mu^2}{s}
+ \bar{{\cal A}}_{y,n} \left( \beta_0^2\,\log^2\frac{\mu^2}{s}
+ \beta_1\, \log\frac{\mu^2}{s}   \right)\bigg)
\nonumber \\ &&
 + {\cal O}(\alpha_s^4)\;.
\label{eq:NNLOmu}
\end{eqnarray}
In here, $s$ denotes the centre-of-mass energy squared and $\mu$ is the
QCD renormalisation scale. The NLO expression is obtained by
suppressing all terms at order $\alpha_s^3$. The first two coefficients
of the QCD $\beta$-function are
\begin{eqnarray}
\beta_0 &=& \frac{11 \CA - 4 T_R \NF}{6}\;,\nonumber  \\
\beta_1 &=& \frac{17 \CA^2 - 10 C_A T_R \NF- 6C_F T_R \NF}{6}\;\;,
\end{eqnarray}
with $C_A=N$, $C_F=(N^2-1)/(2N)$, $T_R=1/2$ for  $N=3$ colours and
$N_F$ quark flavours.

The perturbative
coefficients in (\ref{eq:NNLOmu})
are independent on the centre-of-mass energy. They
are  obtained by integrating parton-level distributions,
 which were calculated recently
to NNLO accuracy~\cite{GehrmannDeRidder:2007hr,GehrmannDeRidder:2007bj,Weinzierl:2009ms}.
These parton-level
calculations are based on a numerical integration of the relevant
three-parton, four-parton and five-parton matrix elements, which are combined
into a parton-level event
generator~\cite{GehrmannDeRidder:2007jk,weinzierlnew,Weinzierl:2009nz} after
subtraction of infrared singular configurations using the antenna subtraction
method~\cite{ourant}. These NNLO event shape
distributions were used subsequently for
improved extractions of the strong coupling
constant~\cite{newas,nnloas,davisonwebber,becherschwartz,asjets},
matched on all-order resummation of logarithmically
enhanced corrections~\cite{becherschwartz,ourresum}, and
used for power correction studies on the thrust
distribution~\cite{davisonwebber}.

The coefficients entering the event shape moments
are computed at a renormalisation scale fixed to
the centre-of-mass energy,
and are therefore just dimensionless numbers for each observable and
each value of $n$. For the first five moments of the
 six event shape variables considered here, they
were computed up to NNLO in~\cite{ourmom,weinzierlmom}.

\section{Dispersive model extended to NNLO}
\label{sec:disp}
Up to now, the dispersive model for power corrections to event shapes
was used in connection with NLO calculations of the perturbative part.
In this context,  one obtains the following, ${1}/{Q}$-dependent power correction \cite{Dokshitzer:1998pt}:
\begin{align}
P=\frac{4C_F}{\pi^2}\cdot\mathcal{M}\cdot\bigg\{\alpha_0-
\bigg[\alpha_s(\mu_R)+\frac{\beta_0}{\pi}\,
\alpha^2_s(\mu_R)\bigg(\ln\frac{\mu_R}{\mu_I}+1+\frac{K}{2\beta_0}\bigg)
+\mathcal{O}(\alpha^3_s)\bigg]\bigg\}\times\frac{\mu_I}{Q}
\end{align}
with the Milan factor $\mathcal{M}=1.49\pm20\%$, which is known at two loops.
Its uncertainty~\cite{dok98} accounts for currently unknown corrections
beyond this loop order. The term in square brackets amounts to the renormalon 
subtraction in the power corrections, expanded to NLO. 

The prediction of the dispersive model can be extended to
match onto the NNLO perturbative prediction, and first steps
in this direction were taken already in~\cite{davisonwebber} for
power corrections to the thrust distribution.

The perturbative ingredients to the dispersive model are the running of the
coupling constant and the relation between the $\overline{{\rm MS}}$-coupling
and the effective coupling, whose definition~\cite{Catani:1990rr}
absorbs universal correction terms from the cusp anomalous dimension.

In the present context, we use
the evolution of the coupling constant to two loops
\begin{equation}
\label{eq:running}
\mu^2 \frac{\d \alpha_s(\mu)}{\d \mu^2} = -\alpha_s(\mu)
\left[\beta_0 \left(\frac{\alpha_s(\mu)}{2\pi}\right)
+ \beta_1 \left(\frac{\alpha_s(\mu)}{2\pi}\right)^2
+ {\cal O}(\alpha_s^3) \right]\,.
\end{equation}
Moreover, the relation between $\overline{{\rm MS}}$-coupling and effective
coupling reads
\begin{eqnarray}
 \alpha^\mathrm{eff}_s&=&\alpha_s\left[1+K\,\frac{\alpha_s}{2\pi}
+L\,\left(\frac{\alpha_s}{2\pi}\right)^2 + {\cal O}(\alpha_s^3) \right]\label{aeffeq}\\
K&=&\bigg(\frac{67}{18}-\frac{\pi^{2}}{6}\bigg)C_A-\frac{5}{9}N_F, \\
L&=&C^2_A\bigg(\frac{245}{24}-\frac{67}{9}\frac{\pi^2}{6}+\frac{11}{6}\zeta_3+\frac{11}{5}\big(\frac{\pi^2}{6}\big)^2\bigg)+C_FN_F\bigg(-\frac{55}{24}+2\zeta_3\bigg)+\nonumber\\
&&\qquad\qquad C_AN_F\bigg(-\frac{209}{108}+\frac{10}{9}\frac{\pi^2}{6}-\frac{7}{3}\zeta_3\bigg)+N^2_F\bigg(-\frac{1}{27}\bigg)\;.
\end{eqnarray}
The coefficient $L$ is obtained from the three-loop cusp anomalous
dimension~\cite{becherneubert,lfactor},
which can be extracted from the three-loop
corrections to the partonic splitting functions~\cite{Moch:2004pa}
 or to the
quark and gluon form factors~\cite{Moch:2005tm}.

The derivation of a generic power correction starts from considering a dimensionless quantity
\begin{equation} F=\int^Q_0\mathrm{d}\mu f(\mu) \label{Feq}\end{equation}
with \begin{equation} f(\mu)\propto a_F\alpha_s(\mu)\frac{\mu^p}{Q^{p+1}}
\end{equation} assuming $F$ to be dimensionless. The value of $p$ determines
the scaling behaviour of the power correction, with $p=0$ for the
leading power correction to event shape variables.

The dispersive model assumes that in the non-perturbative range of
\eqref{Feq} the perturbative strong coupling $\alpha_s(\mu)$ is replaced by an
effective coupling that remains finite for all $\mu$ values. One defines then
the value of the integral over this region by
\begin{equation}
\int^{\mu_I}_0\mathrm{d}\mu\,\alpha_{s,{\rm IR}}(\mu)\,\frac{\mu^p}{Q^{p+1}}
\equiv\frac{\mu_I^{p+1}}{Q^{p+1}(p+1)}{\alpha}_p(\mu_I) \label{a0eq}
\end{equation} introducing an infrared matching scale $\mu_I,
\Lambda_\mathrm{QCD}\ll\mu_I\ll Q$ and ${\alpha}_p$ as a non-perturbative
parameter. One has then to subtract the perturbative part of \eqref{Feq} in the
range from $0$ to $\mu_I$ from the whole integral, that is, the value of
\eqref{a0eq} with $\alpha_{s,\mathrm{IR}}$ replaced by $\alpha_s$.

This perturbative contribution to \eqref{Feq} thus acquires a dependence on the
renormalisation scale $\mu_R$ used in the strong coupling constant. By
requiring $F$ to be scale-independent, one can then infer logarithmic terms
in the non-perturbative contribution to \eqref{Feq}. Applied to the
event-shape power correction $P$ (with $p=0$), this results in
\begin{align}
P&=\frac{4C_F}{\pi^2}\,
{\cal M} \,\bigg\lbrace\alpha_0-\bigg[\alpha_s(\mu_R)+
\frac{\beta_0}{\pi}\,\bigg(1+\ln\left(\frac{\mu_R}{\mu_I}\right)+\frac{K}{2\beta_0}\bigg)\alpha^2_s(\mu_R)+\nonumber
\\
&\qquad
\bigg(2\beta_1\left(1+\ln\left(\frac{\mu_R}{\mu_I}\right)+\frac{L}{2\beta_1}\right)+8\beta^2_0\left(1+\ln\left(\frac{\mu_R}{\mu_I}\right)+\frac{K}{2\beta_0}\right)
\nonumber \\ &\qquad\qquad
+ 4\beta^2_0\ln\left(\frac{\mu_R}{\mu_I}\right)\left(\ln\left(\frac{\mu_R}{\mu_I}\right)+\frac{K}{\beta_0}\right)\bigg)\frac{\alpha^3_s(\mu_R)}{4\pi^2}\bigg]
\bigg\rbrace\times\frac{\mu_I}{Q}\label{peq2}.
\end{align} Together with \eqref{eq:NNLOmu} this gives the full
expression for the event shape observable moments, including
perturbative and non-perturbative contributions.

For $B_T$ and $B_W$ there is a further correction
 to \eqref{peq2}. It arises from the kinematical mismatch between parton 
direction and thrust direction used to define the hemispheres used in the
broadening variables. Retaining (\ref{eq9}), this modification can be accounted 
for by a modification to the power correction. In~\cite{Dokshitzer:1998qp},
this modification was computed to NLO for the first moment as 
\begin{align} P_{\langle
B_W\rangle}&=P\left(\frac{\pi}{\sqrt{8 C_F\hat{\alpha}_s\left(1+\frac{\displaystyle K\hat{\alpha}_s}{\displaystyle 2\pi}\right)}}+\frac{3}{4}-\frac{\beta_0}{6C_F}+\eta_0\right),
\label{bwcorr}\\
P_{\langle
B_T\rangle}&=P\left(\frac{\pi}{\sqrt{4 C_F\hat{\alpha}_s\left(1+\frac{\displaystyle K\hat{\alpha}_s}{\displaystyle 2\pi}\right)}}+\frac{3}{4}-\frac{\beta_0}{3C_F}+\eta_0\right)
\label{btcorr} \end{align}
with $\hat{\alpha}_s(Q)=\alpha_s(e^{-\frac{3}{4}}Q)$ and $\eta_0=-0.6137$.
Corrections to higher moments have not been derived up to now, and we assume 
that they can be approximated by using the above modifications to the 
power correction in all moments. 
The full NNLO expression for these has not been calculated either. 
The potentially
dominant NNLO terms can however be approximated by including the
effective coupling to this order, resulting in
\begin{align}
P_{\langle
B_W\rangle}&=P\left(\frac{\pi}{\sqrt{8 C_F\hat{\alpha}_s\left(1+\frac{\displaystyle K\hat{\alpha}_s}{\displaystyle 2\pi}+\frac{\displaystyle L\hat{\alpha}^2_s}{\displaystyle 4\pi^2}\right)}}+\frac{3}{4}-\frac{\beta_0}{6C_F}+\eta_0\right),
\\ P_{\langle
B_T\rangle}&=P\left(\frac{\pi}{\sqrt{4 C_F\hat{\alpha}_s\left(1+\frac{\displaystyle K\hat{\alpha}_s}{\displaystyle 2\pi}+\frac{\displaystyle L\hat{\alpha}^2_s}{\displaystyle 4\pi^2}\right)}}+\frac{3}{4}-\frac{\beta_0}{3C_F}+\eta_0\right).
\end{align}
However, further NNLO corrections to this expression will reside
in the coefficient $\eta_0$. Therefore, we will treat
$B_W$ and $B_T$
separately from the other variables in the numerical studies in the
following section.

\section{Analysis of JADE and OPAL data}
\label{sec:exp}
The theoretical expressions for event shapes  derived in the previous section
contain two parameters: the strong coupling constant $\alpha_s(M_Z)$ and
the non-perturbative coupling parameter $\alpha_0$.
Using experimental data on event shape moments, it is possible to
fit these parameters.
 The data from the JADE and OPAL experiments~\cite{jadenew}
consists of 18 points at centre-of-mass
 energies between 14.0 and 206.6 GeV for the first five moments of $T$, $C$, $Y_3$, $M_H$, $B_W$ and $B_T$, and have been taken from \cite{pahl}.
For each moment the NLO as well as the NNLO prediction was fitted with $\alpha_s(M_Z)$
and $\alpha_0$ as fit parameters, except for the moments of $Y_3$, which
have no leading $\frac{1}{Q}$ power correction and thus are independent of $\alpha_0$.
For the heavy jet mass, we use only the even moments $\mom{M_H^2}$ and
$\mom{M_H^4}$, since the
theoretical prediction is in terms of $\rho=M_H^2/s$.
\subsection{Fits}
The fits were done using the program ROOT~\cite{root}
 and its $\chi^2$ fit method.
The errors used for the fit were the total errors, composed of the
experimental statistic and systematic errors, added in quadrature.
Based on these, ROOT returned errors on the fit which are displayed
in Tables \ref{table:TfitresultNLO}-\ref{table:BWfitresultNNLO}
in the appendix together with the fit results. For $T$ and $C$
the NNLO values of $\alpha_s(M_Z)$ and $a_0$ seem to be more stable
throughout the moments, as at NNLO they increase less towards higher moments
than at NLO. For $Y_3$ and $\rho$, where the values decrease at higher
moments, this is not the case. These moments show  $\alpha_s(M_Z)$ results
which are significantly lower at NNLO than at NLO.
The $\alpha_s(M_Z)$ values of $B_W$
are much lower than the ones of the other observables, and do not change much
from NLO to NNLO. For $B_T$ the $\alpha_s(M_Z)$ values at NNLO are lower than at NLO.
Both are exceptionally stable throughout the different moments. The
$\alpha_0$ values of all moments are higher at NNLO than at NLO.

\begin{figure}[t]
\centering
\includegraphics[width=14cm]{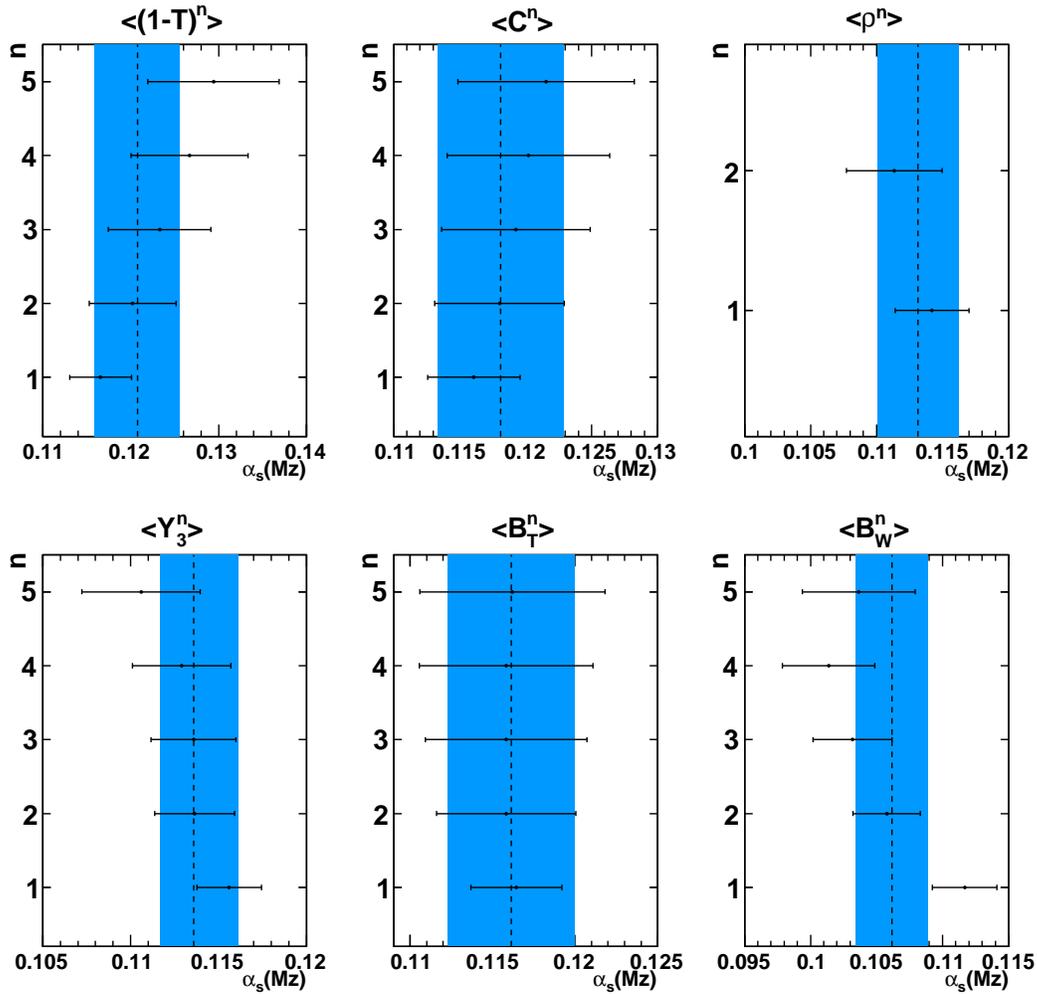}
\caption[Error band plot of the individual measurements for $\alpha_s(M_Z)$]{Plot of the individual measurements for $\alpha_s(M_Z)$. The shaded region corresponds to the error band defined by the weighted mean for the event shape and the total error on it.}
\label{fig:as}
\end{figure}
\begin{figure}[t]
\centering
\includegraphics[width=14cm]{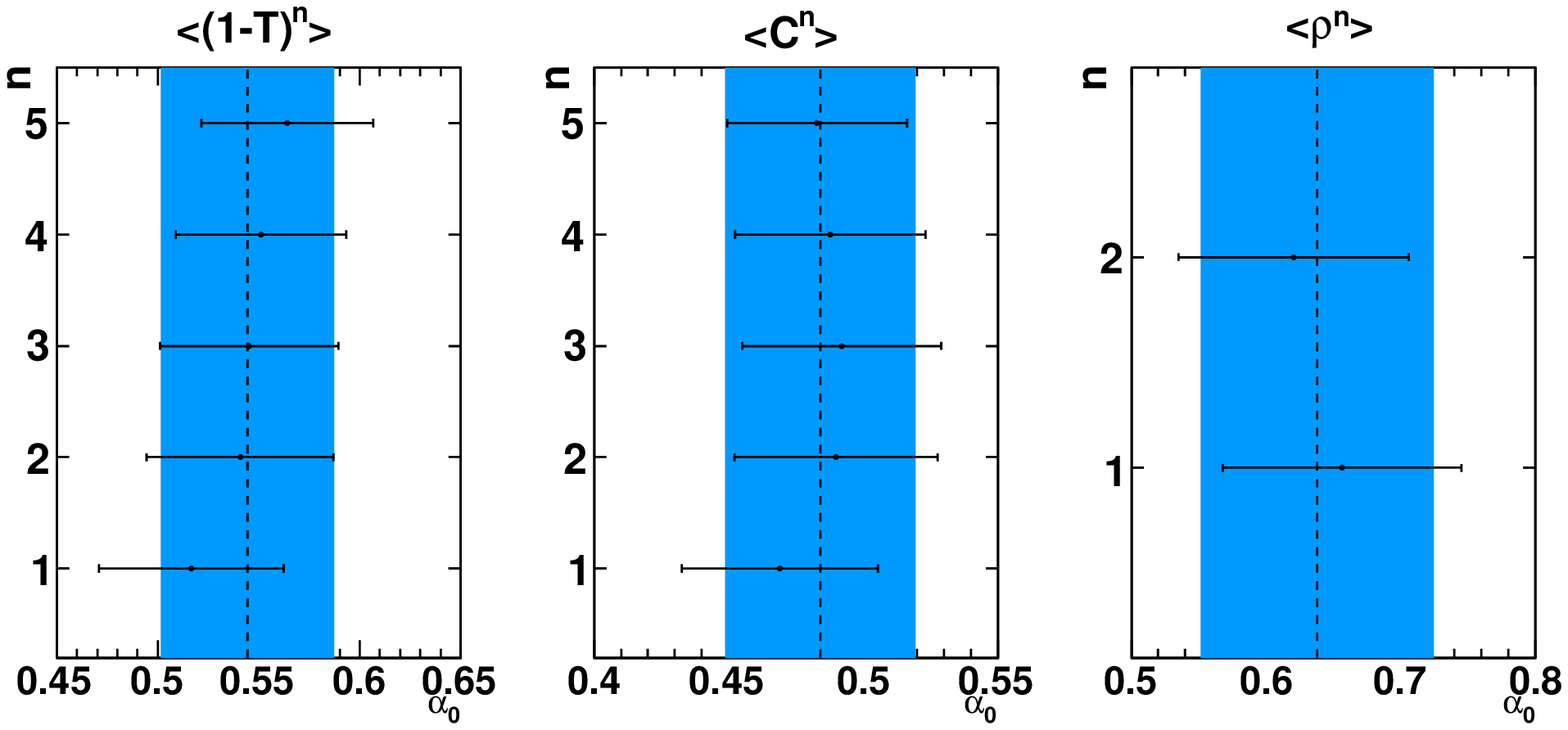}%
\\
\includegraphics[width=9.3cm]{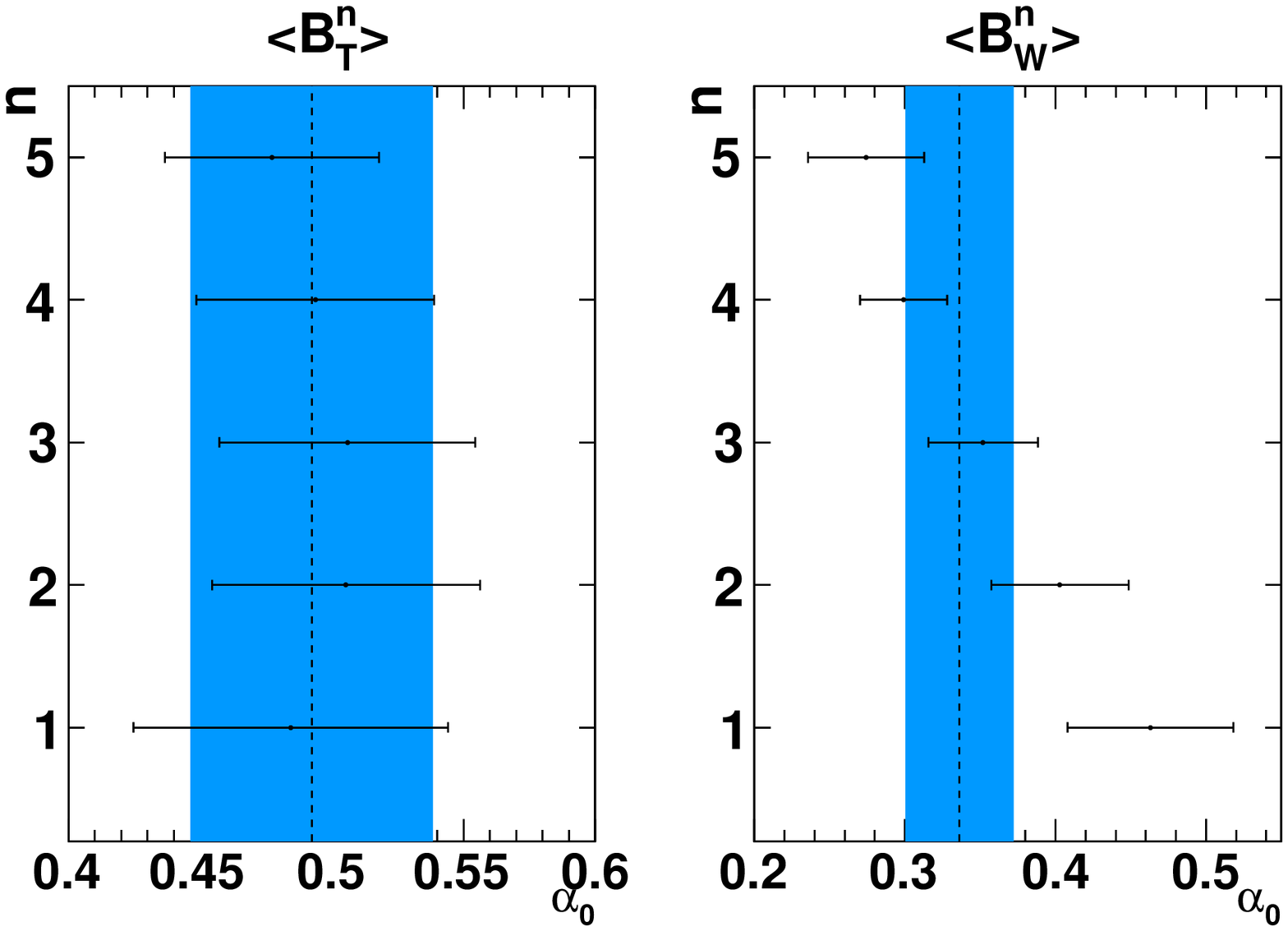}%
\caption{Error band plot of the individual measurements for $\alpha_0$}
\label{fig:a0}
\end{figure}

\subsection{Theoretical systematic errors}
There are different parameters in the theoretical prediction which may influence the results displayed above, namely the matching scale $\mu_I$, the renormalisation scale $\mu_R$ and the Milan factor $\mathcal{M}$. In order to estimate the resulting theoretical uncertainty on $\alpha_s(M_Z)$ and $\alpha_0$, the fits were repeated, $\mu_I$, $\mu_R$ and $\mathcal{M}$ being separately varied by a certain amount.
\begin{table}[t]
\begin{center}
\begin{tabular}{llll} \hline
&nominal value &  up variation &  down variation \\ \hline
$\mu_I$[GeV] & 2 & 3 & 1 \\ \hline
$x_\mu$ & 1 & 2 & 0.5 \\ \hline
$\mathcal{M}$ & 1.49 & 1.788 (+20\%) & 1.192 (-20\%) \\ \hline
\end{tabular}
\end{center}
\caption{Table of the $\mu_I$, $x_\mu$ and $\mathcal{M}$ variations}\label{varTab}
\end{table}

For this purpose the scaling factor $x_\mu = \frac{\mu_R}{Q}$ was introduced. The uncertainty on
 the corresponding parameter was then taken to  be the difference between the nominal and the new value returned by ROOT. In order to get a total systematic error, the greater
values of the up and down uncertainties were determined and quadratically added. As $\alpha_0$ depends directly on $\mu_I$ no error was determined for 
this variation. For $Y_3$ there is only an error on $\alpha_s(M_Z)$ coming from the $x_\mu$ variation, since the theoretical 
description of this observable does not contain a contribution from the leading power correction, and is thus independent on  $\mu_I$ 
and $\mathcal{M}$. At NLO the fit to the moment $\mom{C^3}$ suffers from a numerical instability by scaling up $\mathcal{M}$ by $20\%$. The numbers reported in Table~\ref{table:CfitresultNLO} refer to an up variation of $19\%$.

The NLO error on $\alpha_s(M_Z)$ agrees well with
the values of \cite{jadenew}. At NNLO, it is reduced by more than
half throughout all event shape observables except $B_W$, confirming a good description by the NNLO prediction.
Unfortunately, this is not the case for the error on $\alpha_0$. It does not change much from NLO to NNLO, even increasing a little in the first moments due to the higher $x_\mu$ uncertainty at NNLO and decreasing slightly at the higher moments, with exception, again, of $B_W$.
Analysing the different sources of the systematical errors, we observe that
the error on $\alpha_s(M_Z)$ is clearly dominated by the $x_{\mu}$ variation,
while the largest contribution to the error on $\alpha_0$ comes
from the uncertainty on the Milan factor ${\cal M}$. Since this
uncertainty has not been improved in the current study, it is understandable
that the systematic error on  $\alpha_0$ remains unchanged. This finding
clearly motivates the need for a three-loop calculation of the Milan factor.
However, it is
very important to note that the uncertainty on the Milan factor
has little impact on the extraction of  $\alpha_s(M_Z)$, thereby
demonstrating the systematic decoupling of perturbative and non-perturbative
effects in the dispersive model.
\begin{table}[t!]
\centering
\begin{tabular}{{|l|c|cc|c|}}
\hline
\multicolumn{5}{|c|}{\bf{NNLO}}\\\hline Observable &$\alpha_s\left(M_Z\right)$ & Experimental Error & Theoretical Error & Total Error\\\hline 
$\tau$                 & 0.1208 & 0.0018 & 0.0045 & 0.0048\\ 
C                      & 0.1181 & 0.0013 & 0.0046 & 0.0048\\ 
$\rho$                 & 0.1131 & 0.0024 & 0.0019 & 0.0031\\ 
$Y_3$                  & 0.1139 & 0.0016 & 0.0015 & 0.0022\\ 
$B_T$                  & 0.1161 & 0.0014 & 0.0036 & 0.0038\\ 
$B_W$                  & 0.1062 & 0.0021 & 0.0018 & 0.0027\\\hline\hline 
Total                  & 0.1131 & 0.0017 & 0.0022 & 0.0028\\ 
Total w/o $B_T$,$B_W$  & 0.1153 & 0.0017 & 0.0023 & 0.0028\\\hline\hline
 Observable &$\alpha_0$& Experimental Error & Theoretical Error & Total Error\\\hline 
$\tau$                 & 0.5444 & 0.0184 & 0.0388 & 0.0430\\ 
C                      & 0.4841 & 0.0066 & 0.0347 & 0.0353\\ 
$\rho$                 & 0.6380 & 0.0270 & 0.0824 & 0.0867\\ 
$Y_3$                  & - & - & - & -\\ 
$B_T$                  & 0.4924 & 0.0102 & 0.0449 & 0.0460\\ 
$B_W$                  & 0.3362 & 0.0125 & 0.0338 & 0.0360\\\hline\hline 
Total                  & 0.4604 & 0.0108 & 0.0359 & 0.0375\\
Total w/o $B_T$,$B_W$  & 0.5132 & 0.0115 & 0.0381 & 0.0398\\\hline
\end{tabular}
\caption{Table of the $\alpha_s(M_Z)$ and $\alpha_0$ results for the individual moments and the global weighted average.}
\label{tab:cNNLO}
\end{table}

For the higher moments ($n\geq 2$) of the jet broadenings $B_W$ and $B_T$, 
the kinematical modifications to the power correction are not known at present.
We have approximated them in the above fits
by the corrections to the first moments, 
given to NLO and NNLO in the previous section. If we do not apply these 
correction to the higher moments, the mutual consistency of the parameter 
extractions from different moments of $B_T$ deteriorates considerably, 
while only minor improvements in consistency are observed on $B_W$. 

Including empirical hadron mass corrections~\cite{Salam:2001bd,salamp} 
from PYTHIA affects in particular the parameter extraction from $\rho$, resulting in 
values of  $\alpha_s(M_Z)$ and $\alpha_0$  from $\rho$ much lower than from 
the other variables. Since these corrections may interplay with other 
non-perturbative parameters in PYTHIA, we do not include them in our default 
fits or error estimates.

By taking the weighted means over the corresponding values 
from all moments of all observables one gets combined values for 
 $\alpha_s(M_Z)$ and $\alpha_0$. The weights are given by the inverse of the total error squared and are normalized such that the sum over all weights is equal to one. For the errors one has to take care of the correlation between the errors of the single measurements. The correlation matrix for $\alpha_s(M_Z)$ and $\alpha_0$ is in first approximation equal to the correlation matrix for the event shape moments, since  the variable transformation is
linear in first approximation.
The correlation matrix for the event shape moments  is given in \cite{pahl}.
We first combine the measurements from different moments of the same 
observable. Figures~\ref{fig:as} and~\ref{fig:a0} compare the combined NNLO 
results on the $\alpha_s(M_Z)$  and $\alpha_0$ measurements. Owing to 
the large correlation between individual moments of the same observable, the 
combined errors are only marginally smaller than the errors obtained
from single measurements. The combined results and their errors are summarised 
in Table~\ref{tab:cNNLO}. From this Table, we clearly observe that the 
the theoretical error on the extraction of $\alpha_S(M_Z)$ 
from $\rho$, $Y_3$ and $B_W$ 
is considerably smaller than from $\tau$, $C$ and $B_T$. It was observed 
previously in~\cite{ourmom} that the moments of the former three shape 
variables receive moderate NNLO corrections for all $n$, while the NNLO 
corrections for the latter three are large already for $n=1$ and 
increase with $n$. Consequently, 
the theoretical description of the moments of $\rho$, $Y_3$ and $B_W$ 
displays a higher perturbative stability, which is reflected in the 
theoretical uncertainty on  $\alpha_S(M_Z)$ derived from them. 

In a second step, we combine the $\alpha_s(M_Z)$ and $\alpha_0$ measurements 
obtained from different event shape variables. 
Taking the weighted mean over all values, but  excluding the values for
the moments of $B_W$ and $B_T$ where the theoretical description
is incomplete,  we obtain at NNLO:
\begin{align}
\alpha_s(M_Z)&=0.1153\pm0.0017(\mathrm{exp})\pm0.0023(\mathrm{th}),\nonumber \\
\alpha_0&=0.5132\pm0.0115(\mathrm{exp})\pm0.0381(\mathrm{th})\,,
\label{eq:final}
\end{align}
where the errors have been derived taking into account the correlation between the moments of different event shapes. Including the values for $B_W$ and $B_T$ modifies this result to:
\begin{align}
\alpha^B_s(M_Z)&=0.1131\pm0.0017(\mathrm{exp})\pm0.0022(\mathrm{th}), \nonumber\\
\alpha^B_0&=0.4604\pm0.0108(\mathrm{exp})\pm0.0359(\mathrm{th})\,. \nonumber 
\end{align}
These latter values are however quoted only to illustrate the impact of
including the broadenings. The default fit result is (\ref{eq:final}), where
only observables with a consistent theoretical description are included.

\begin{figure}[t]
\centering
\includegraphics[width=5.0cm]{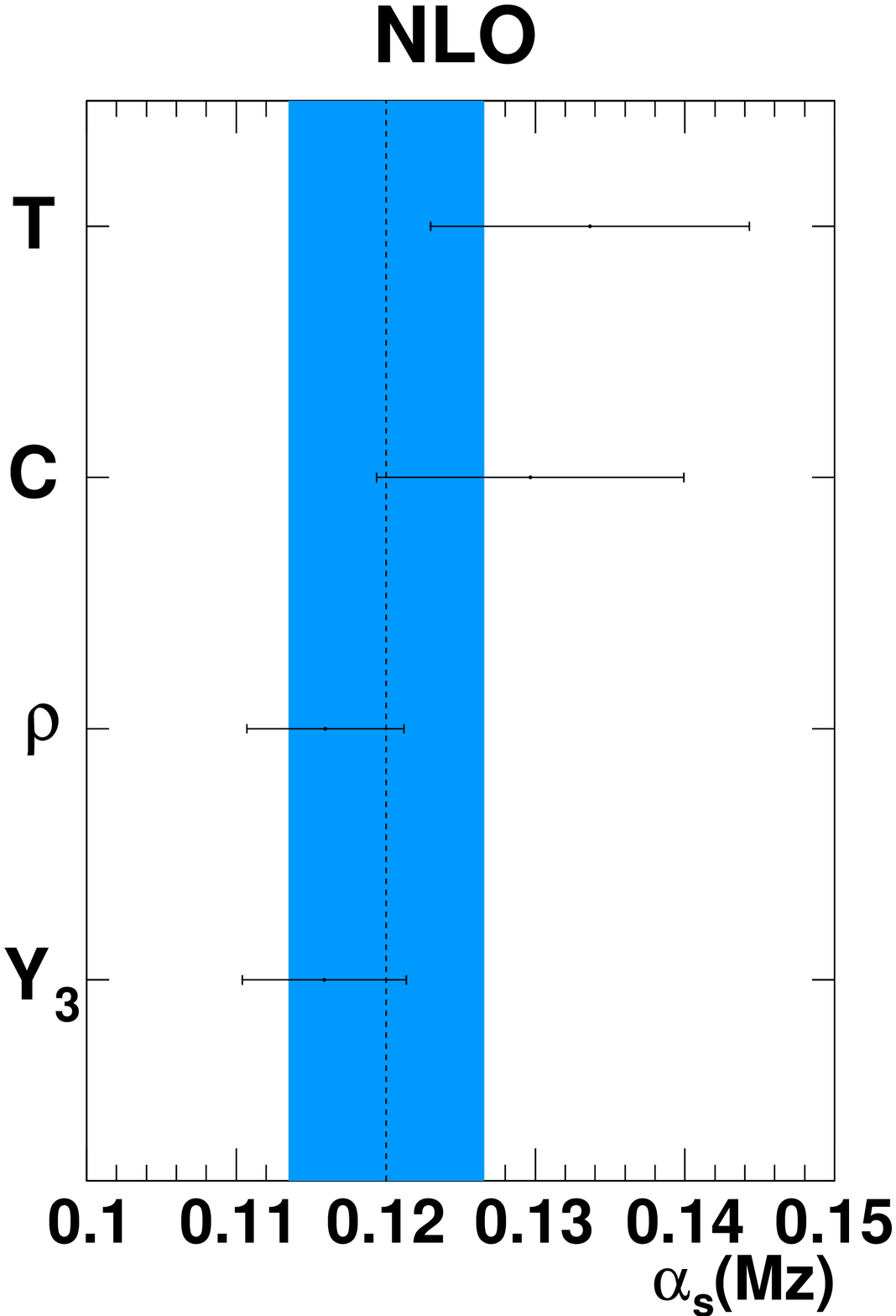}%
\qquad\qquad
\includegraphics[width=5.0cm]{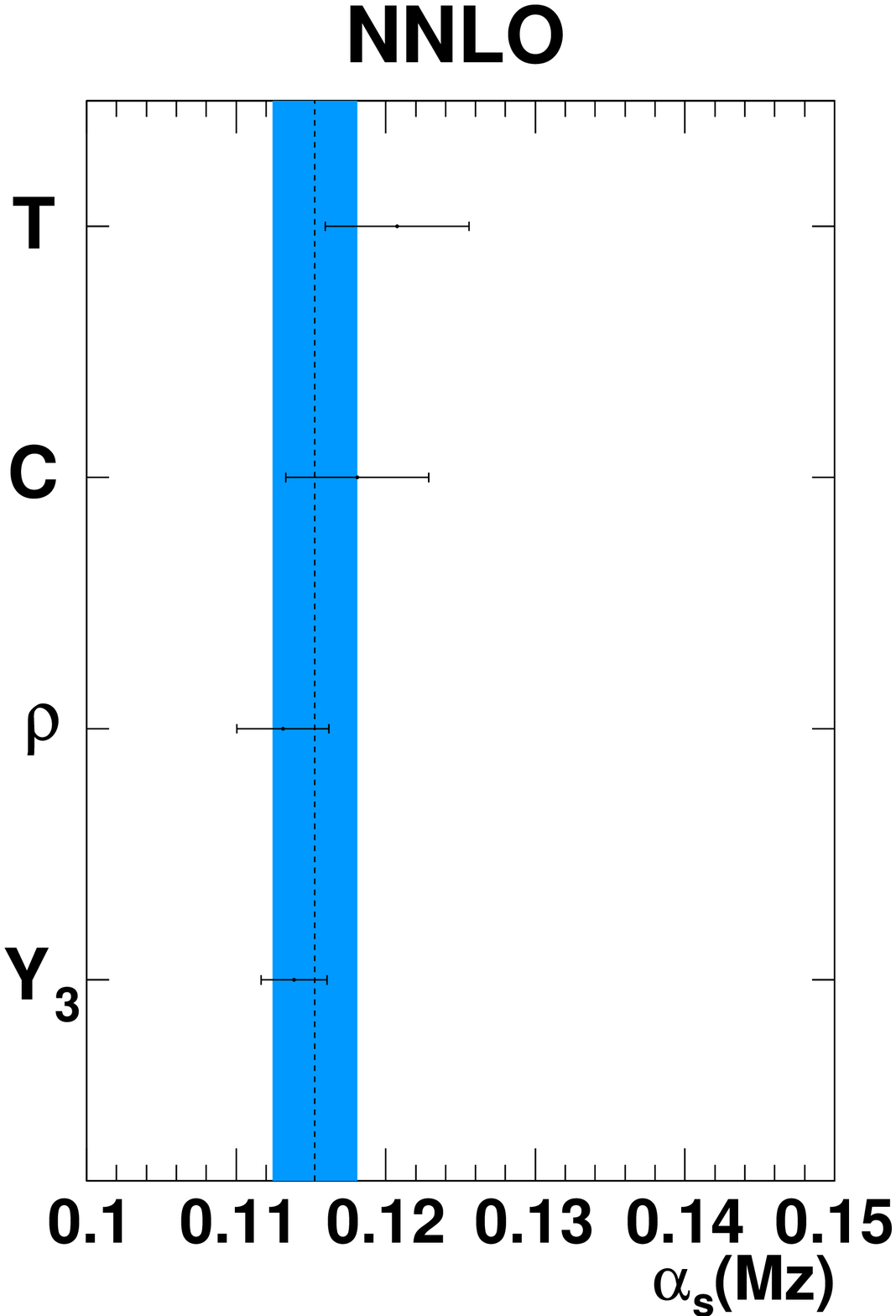}%
\\
\includegraphics[width=5.0cm]{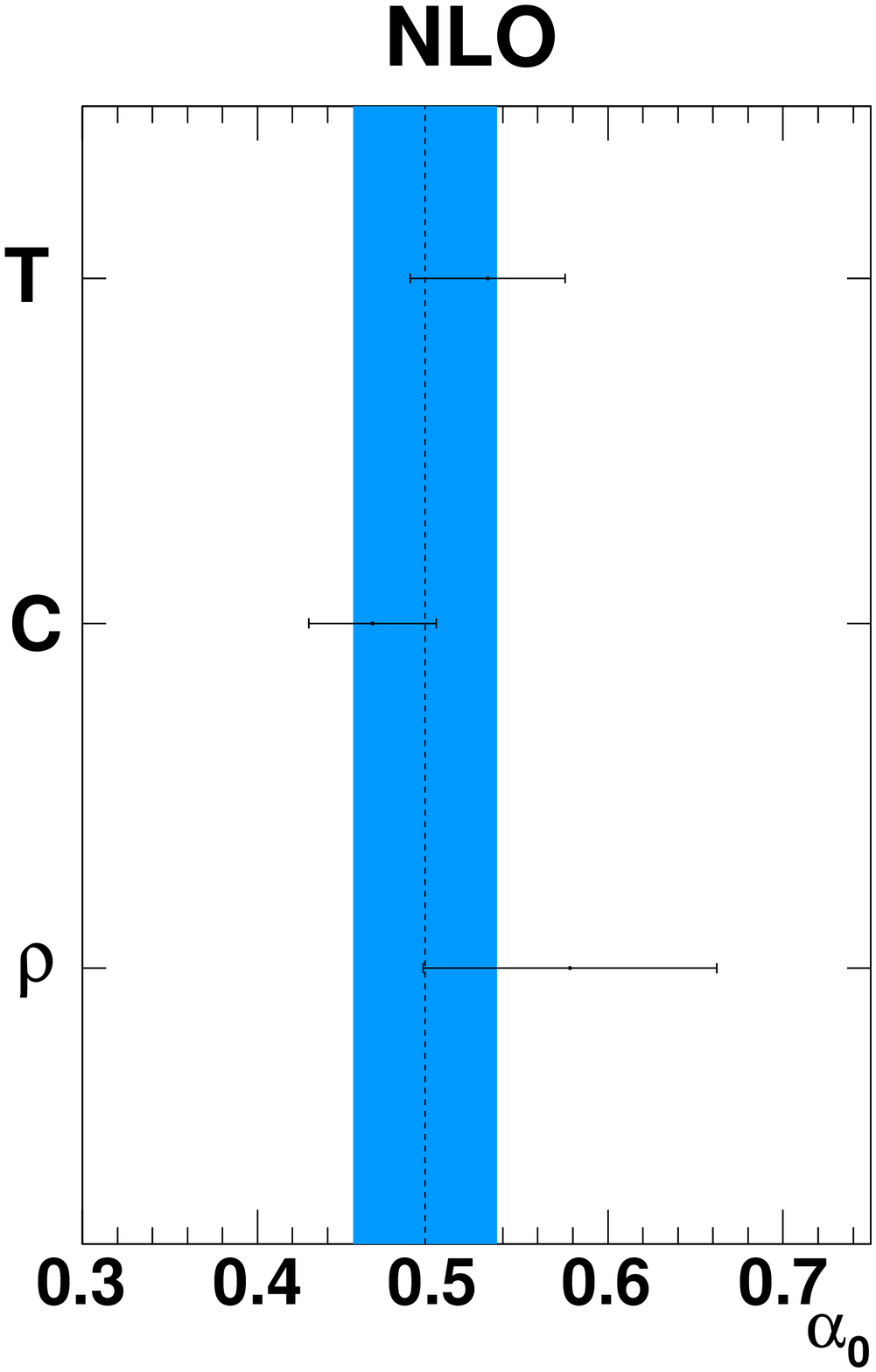}%
\qquad\qquad
\includegraphics[width=5.0cm]{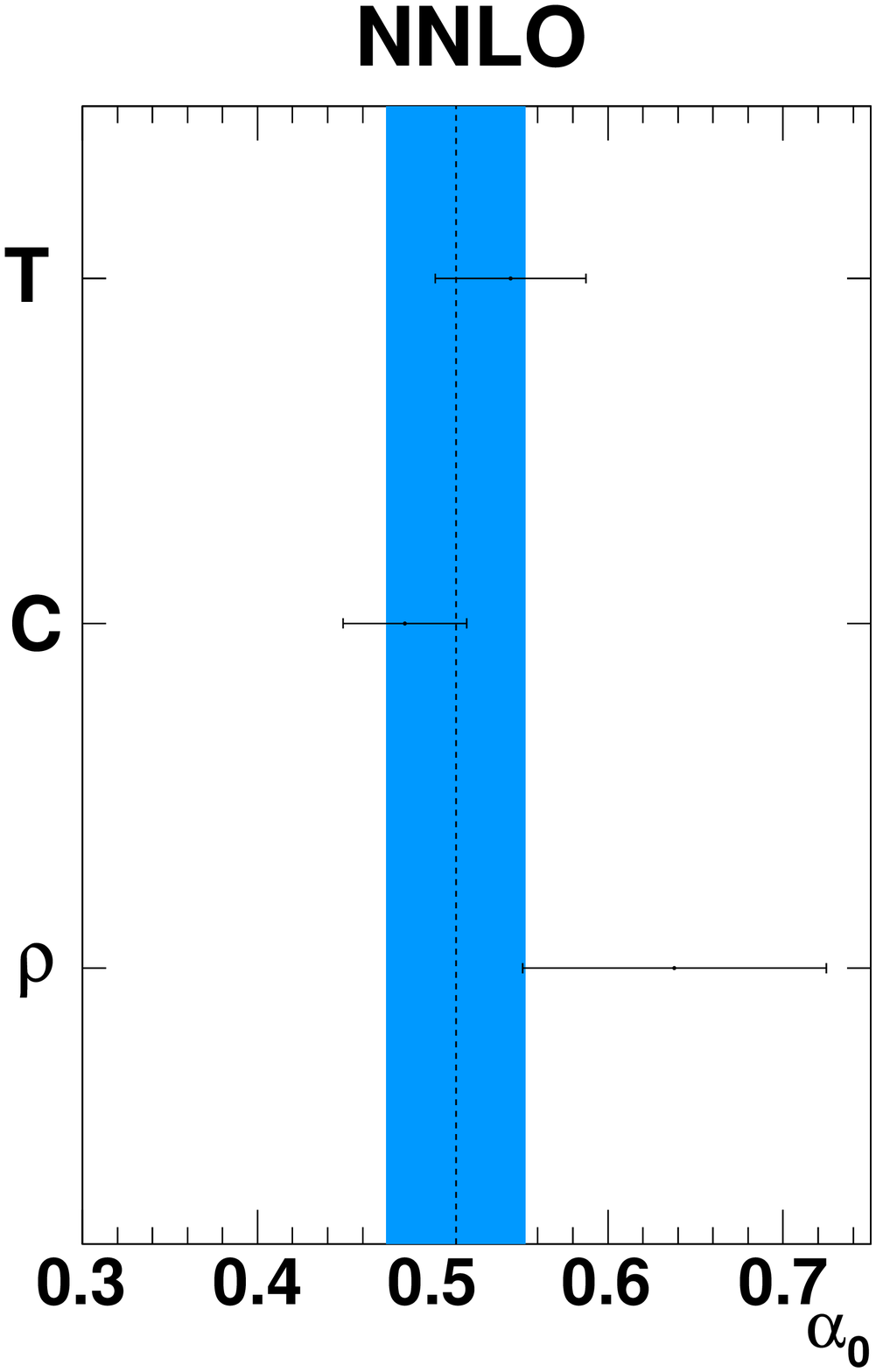}%
\caption[Error band plot of the final results]{Error band plot of the final results. The points for $\alpha_s(M_Z)$ are $C$, $T$, $Y_3$, $M_H$ and for $\alpha_0$ $C$, $T$, $M_H$.}
\label{fig:final}
\end{figure}

To illustrate the improvement due to the inclusion of the NNLO 
corrections, we also quote the corresponding NLO results. Based on 
$\tau$, $C$, $\rho$ and $Y_3$, we obtain:
\begin{align}
\alpha^{{\rm NLO}}_s(M_Z)&=0.1200\pm0.0021(\mathrm{exp})\pm0.0062(\mathrm{th}),\nonumber \\
\alpha^{{\rm NLO}}_0&=0.4957\pm0.0118(\mathrm{exp})\pm0.0393(\mathrm{th})\,, \nonumber 
\end{align}
while inclusion of $B_W$ and $B_T$ modifies this to 
\begin{align}
\alpha^{{\rm NLO},B}_s(M_Z)&=0.1147\pm0.0020(\mathrm{exp})\pm0.0046(\mathrm{th}),\nonumber \\
\alpha^{{\rm NLO}, B}_0&=0.4019\pm0.0130(\mathrm{exp})\pm0.0296(\mathrm{th})\,, \nonumber 
\end{align}
We compare the NLO and NNLO combinations in Figure~\ref{fig:final}. It can be 
seen very clearly that the measurements obtained from the different variables 
are consistent with each other within errors. The average of $\alpha_s(M_Z)$ is 
dominated by the measurements based on $\rho$ and $Y_3$, which have the 
smallest theoretical uncertainties. From NLO to NNLO, the error on 
 $\alpha_s(M_Z)$  is reduced by a factor two, and the result shifts towards 
the lower end of the NLO error band, as was already 
the case in the individual measurements. No improvement and no shift in the 
central value 
between NLO and NNLO is seen on  
$\alpha_0$.  

\section{Comparison with PYTHIA hadronization corrections}
\label{sec:MCcomparison}
The primary motivation for studying power corrections to 
moments of event shapes in the dispersive model comes from the observation 
that the commonly used method to derive hadronization corrections 
 from multi-purpose event generator programs may be unreliable~\cite{newas}.
\begin{figure}[t]
\centering
\includegraphics[width=12.0cm]{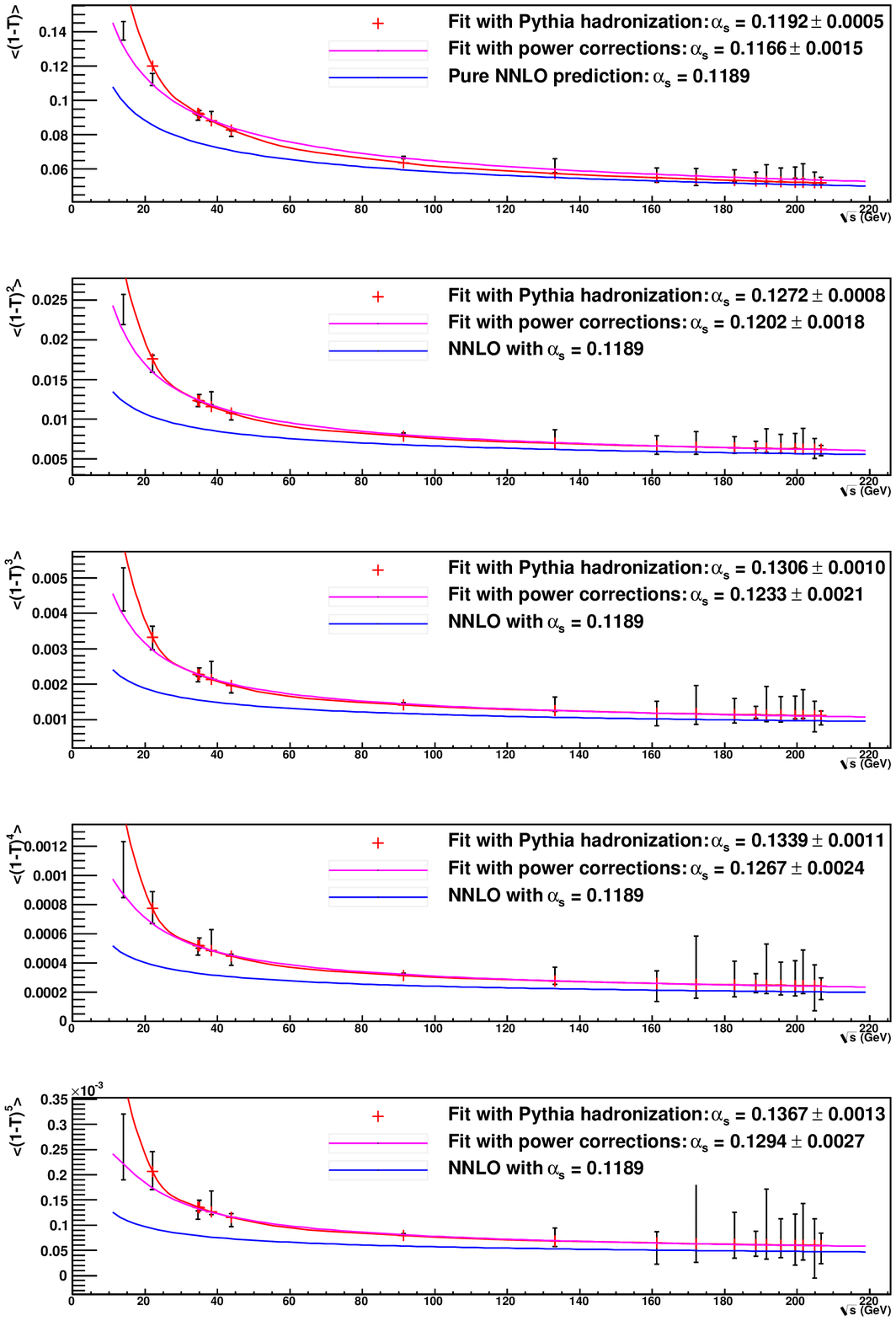}%
\caption{Comparison of fits with hadronization corrections from 
PYTHIA and power corrections from the dispersive model.}
\label{fig:tfit}
\end{figure}

To quantify the difference of both approaches to hadronization corrections, 
we compare them on the example of the moments of $1-T$. For this comparison, 
we extracted the PYTHIA~\cite{Sjostrand:2000wi}
 hadronization corrections to these moments 
from the ratio of PYTHIA hadron level and parton level results. Using these 
corrections in combination with the NNLO perturbative expressions for the 
event shape moments, we repeated the fit of $\alpha_S(M_Z)$ on the 
different moments of $1-T$. The results are displayed 
and compared with the fits in the dispersive model in Figure~\ref{fig:tfit}. 
We observe that both approaches yield a reasonable description of the
experimental data, but that 
the resulting values of $\alpha_s(M_Z)$ are considerably 
larger when applying hadronization corrections extracted from PYTHIA.
 Given that the perturbative contribution increases 
monotonously with $\alpha_s(M_Z)$, this indicates that 
the hadronization corrections in PYTHIA are considerably smaller (and 
perhaps underestimated) than 
those obtained in the dispersive model.
\begin{figure}[t]
\centering
\includegraphics[width=10.0cm]{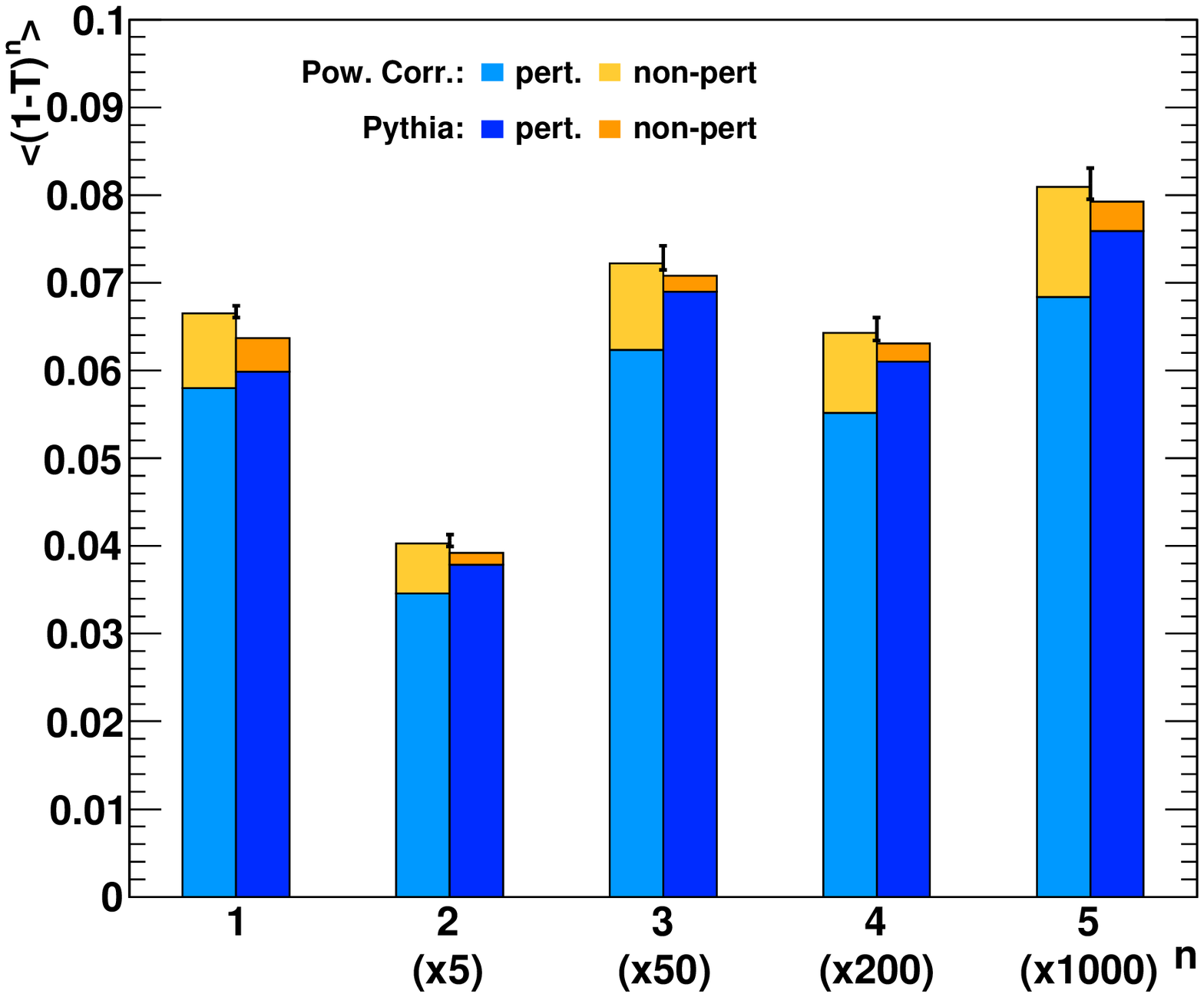}%
\caption{Perturbative and non-perturbative contributions to the moments of 
$1-T$ at $\sqrt{s}=M_Z$ as predicted by power corrections (left) and PYTHIA (right).}
\label{fig:tcorr}
\end{figure}

This observation is quantified on the moments of $(1-T)$
at $\sqrt{s}=M_Z$, displayed in Figure~\ref{fig:tcorr}. Depending on the 
moment number, we observe that the PYTHIA hadronization corrections are 
between two and four times smaller than those obtained from the 
dispersive model. It can also be seen that 
the PYTHIA-based  predictions are systematically below the experimental data,
which perhaps indicates that, despite the decent agreement on the full 
range of energies, Figure~\ref{fig:tfit}, PYTHIA fails in the precise 
description of the energy dependence of the hadronization corrections. 

Our comparison suggests strongly that hadronization corrections extracted 
from PYTHIA (or from other comparable multi-purpose event generator 
programs~\cite{herwigpythia}) are lower than power corrections 
obtained from analytical hadronization models. This can be partially understood from the fact that perturbative predictions of PYTHIA are larger than fixed order calculations since they are based on a parton-shower approximation to the all order resummation result, which is known to shift the distribution away from the two jet singularity. As a consequence, 
using PYTHIA hadronization corrections to analyse data in view of 
precision extractions of $\alpha_s(M_Z)$ may result in anomalously large 
values, since the missing numerical magnitude of the power corrections must
be compensated by a larger perturbative contribution. 

\section{Conclusions}

In this paper, we studied the perturbative and non-perturbative
contributions to the moments of event shapes in $e^+e^-$ annihilation.
In view of the recently calculated NNLO perturbative
contributions~\cite{ourmom,weinzierlmom}  to the
event shape moments, we extended the dispersive model for non-perturbative
power corrections~\cite{Dokshitzer:1995qm,Dokshitzer:1997ew,Dokshitzer:1998pt}
 to include all logarithmic corrections to this order. The normalisation
of the power correction (the Milan factor \cite{Dokshitzer:1998pt})
is however still restricted
to NLO accuracy, and specific corrections~\cite{Dokshitzer:1998qp}
 to the jet broadenings  $B_W$ and $B_T$ are also only included to NLO.

We used this newly
obtained theoretical description of the event shape moments
to reanalyse data from JADE and OPAL in view of a determination of the
strong coupling constant $\alpha_s(M_Z)$ and of the non-perturbative
parameter $\alpha_0$. We observed that inclusion of the NNLO corrections
results in a considerably
better consistency among the values extracted from different
moments  of the same variable, and an improved consistency among the different
variables. Averaging over the different moments and different shapes
(excluding $B_W$ and $B_T$, where the theoretical description is incomplete,
and taking proper account of the uncertainty due to missing terms in the Milan factor), we obtain the following combined values:
\begin{align}
\alpha_s(M_Z)&=0.1153\pm0.0017(\mathrm{fit})\pm0.0023(\mathrm{th}),\nonumber \\
\alpha_0&=0.5132\pm0.0115(\mathrm{fit})\pm0.0381(\mathrm{th})\,,\nonumber 
\end{align}

Compared to previous NLO results, the theoretical error on $\alpha_S(M_Z)$
(which is dominated by the scale variation, improved at NNLO)
is reduced by a factor of two, while the error on $\alpha_0$
(which is dominated by the uncertainty on the Milan factor) remains unchanged.
We observed that the sources of uncertainty on  $\alpha_S(M_Z)$ and  $\alpha_0$
largely decouple. An improvement on $\alpha_0$ will only be achievable
once the three-loop corrections to the Milan factor become available.

It is noteworthy that
application of the  dispersive model to hadronization corrections
results in a considerably lower value of $\alpha_s(M_Z)$ from event shapes
than pervious studies based on Monte Carlo hadronization models~\cite{newas},
and in better agreement with measurements from other observables~\cite{pdg}.
A direct comparison hints to an underestimation of hadronization effects in 
the Monte Carlo models.
This feature  has been observed previously also on the thrust
distribution~\cite{davisonwebber}. Revisiting the hadronization models
in multi-purpose Monte Carlo programs appears to be mandatory for
meaningful precision QCD studies at colliders.

\section*{Acknowledgements}
We would like to thank Hasko Stenzel, Gudrun Heinrich and Gavin Salam
for useful discussions.
This research was supported by the Swiss National Science Foundation
(SNF) under contract 200020-126691.

\begin{appendix}

\section{Tables of results}
In this appendix, we collect the extractions of $\alpha_s(M_Z)$ and 
$\alpha_0$ at NLO and NNLO from individual moments of the six event shape 
variables: $\tau$, $C$, $\rho$, $Y_3$, $B_T$, $B_W$. 
\begin{table}[h!]
\centering
\begin{tabular}{{l|ccccc}}
NLO &$\left<\tau\right>$ & $\left<\tau^2\right>$ & $\left<\tau^3\right>$ & $\left<\tau^4\right>$ & $\left<\tau^5\right>$\\\hline\hline 
$\chi^{2}/\mathrm{dof}$                       & 1.0043 & 1.0565 & 0.8399 & 0.6459 & 0.4740 \\\hline 
$\alpha_s(M_Z)$                               & 0.1242 & 0.1344 & 0.1416 & 0.1479 & 0.1534 \\\hline 
Experimental Error                            & 0.0018 & 0.0023 & 0.0028 & 0.0033 & 0.0038 \\\hline 
$x_{\mu}$ variation: $x_{\mu}=0.5$            & -0.0054 & -0.0083 & -0.0101 & -0.0115 & -0.0129 \\ 
\hspace{2cm}         $x_{\mu}=2.0$            & 0.0066 & 0.0102 & 0.0123 & 0.0143 & 0.0162 \\ 
$\mu_{I}$ variation: $\mu_{I}=1.0$ Gev        & 0.0025 & 0.0035 & 0.0038 & 0.0045 & 0.0054 \\ 
\hspace{1.99cm}      $\mu_{I}=3.0$ Gev        & -0.0019 & -0.0025 & -0.0027 & -0.0031 & -0.0037 \\ 
$\mathcal{M}$ variation: $\mathcal{M}-20\,\%$ & 0.0012 & 0.0016 & 0.0017 & 0.0020 & 0.0024 \\ 
\hspace{2.02cm}          $\mathcal{M}+20\,\%$ & -0.0011 & -0.0014 & -0.0016 & -0.0018 & -0.0021 \\\hline 
Theoretical Error                             & 0.0072 & 0.0109 & 0.0130 & 0.0151 & 0.0172 \\\hline\hline 
$\alpha_0$                                    & 0.4782 & 0.5147 & 0.5359 & 0.5521 & 0.5744 \\\hline 
Experimental Error                            & 0.0151 & 0.0152 & 0.0189 & 0.0222 & 0.0243 \\\hline 
$x_{\mu}$ variation: $x_{\mu}=0.5$            & 0.0017 & 0.0004 & -0.0038 & -0.0065 & -0.0081 \\ 
\hspace{2cm}         $x_{\mu}=2.0$            & -0.0000 & -0.0001 & 0.0030 & 0.0051 & 0.0064 \\ 
$\mathcal{M}$ variation: $\mathcal{M}-20\,\%$ & 0.0432 & 0.0423 & 0.0405 & 0.0377 & 0.0375 \\ 
\hspace{2.02cm}          $\mathcal{M}+20\,\%$ & -0.0306 & -0.0307 & -0.0298 & -0.0284 & -0.0290 \\\hline 
Theoretical Error                             & 0.0433 & 0.0423 & 0.0406 & 0.0382 & 0.0384 \\\hline 
\end{tabular}
\caption{Results for $\alpha_{s}(Q)$ and $\alpha_{0}$ as obtained from fits to $\tau$ moments measured by JADE and OPAL for centre-of-mass energies between 14.0 and 206.6 GeV using theoretical NLO predictions.} 
\label{table:TfitresultNLO} 
\end{table}

\begin{table}[h!]
\centering
\begin{tabular}{{l|ccccc}}
NNLO &$\left<\tau\right>$ & $\left<\tau^2\right>$ & $\left<\tau^3\right>$ & $\left<\tau^4\right>$ & $\left<\tau^5\right>$\\\hline\hline 
$\chi^{2}/\mathrm{dof}$                       & 0.9889 & 0.9411 & 0.7284 & 0.5526 & 0.3997 \\\hline 
$\alpha_s(M_Z)$                               & 0.1166 & 0.1202 & 0.1233 & 0.1267 & 0.1294 \\\hline 
Experimental Error                            & 0.0015 & 0.0018 & 0.0021 & 0.0024 & 0.0027 \\\hline 
$x_{\mu}$ variation: $x_{\mu}=0.5$            & -0.0020 & -0.0034 & -0.0042 & -0.0048 & -0.0054 \\ 
\hspace{2cm}         $x_{\mu}=2.0$            & 0.0025 & 0.0042 & 0.0051 & 0.0058 & 0.0065 \\ 
$\mu_{I}$ variation: $\mu_{I}=1.0$ Gev        & 0.0017 & 0.0017 & 0.0017 & 0.0019 & 0.0022 \\ 
\hspace{1.99cm}      $\mu_{I}=3.0$ Gev        & -0.0011 & -0.0011 & -0.0011 & -0.0013 & -0.0014 \\ 
$\mathcal{M}$ variation: $\mathcal{M}-20\,\%$ & 0.0009 & 0.0010 & 0.0009 & 0.0011 & 0.0012 \\ 
\hspace{2.02cm}          $\mathcal{M}+20\,\%$ & -0.0009 & -0.0009 & -0.0009 & -0.0010 & -0.0011 \\\hline 
Theoretical Error                             & 0.0032 & 0.0046 & 0.0054 & 0.0062 & 0.0070 \\\hline\hline 
$\alpha_0$                                    & 0.5165 & 0.5408 & 0.5452 & 0.5512 & 0.5641 \\\hline 
Experimental Error                            & 0.0135 & 0.0152 & 0.0194 & 0.0223 & 0.0246 \\\hline 
$x_{\mu}$ variation: $x_{\mu}=0.5$            & 0.0140 & 0.0075 & 0.0016 & -0.0008 & -0.0023 \\ 
\hspace{2cm}         $x_{\mu}=2.0$            & -0.0078 & -0.0045 & -0.0001 & 0.0019 & 0.0031 \\ 
$\mathcal{M}$ variation: $\mathcal{M}-20\,\%$ & 0.0415 & 0.0430 & 0.0396 & 0.0357 & 0.0347 \\ 
\hspace{2.02cm}          $\mathcal{M}+20\,\%$ & -0.0298 & -0.0308 & -0.0286 & -0.0264 & -0.0261 \\\hline 
Theoretical Error                             & 0.0438 & 0.0436 & 0.0397 & 0.0358 & 0.0348 \\\hline 
\end{tabular}
\caption{Results for $\alpha_{s}(Q)$ and $\alpha_{0}$ as obtained from fits to $\tau$ moments measured by JADE and OPAL for centre-of-mass energies between 14.0 and 206.6 GeV using theoretical NNLO predictions.} 
\label{table:TfitresultNNLO} 
\end{table}

\begin{table}[h!]
\centering
\begin{tabular}{{l|ccccc}}
NLO &$\left<C\right>$ & $\left<C^2\right>$ & $\left<C^3\right>$ & $\left<C^4\right>$ & $\left<C^5\right>$\\\hline\hline 
$\chi^{2}/\mathrm{dof}$                       & 1.1849 & 1.5245 & 1.5651 & 1.5446 & 1.4094 \\\hline 
$\alpha_s(M_Z)$                               & 0.1230 & 0.1308 & 0.1347 & 0.1374 & 0.1407 \\\hline 
Experimental Error                            & 0.0013 & 0.0016 & 0.0020 & 0.0023 & 0.0026 \\\hline 
$x_{\mu}$ variation: $x_{\mu}=0.5$            & -0.0052 & -0.0079 & -0.0091 & -0.0100 & -0.0108 \\ 
\hspace{2cm}         $x_{\mu}=2.0$            & 0.0063 & 0.0096 & 0.0111 & 0.0122 & 0.0134 \\ 
$\mu_{I}$ variation: $\mu_{I}=1.0$ Gev        & 0.0029 & 0.0045 & 0.0051 & 0.0057 & 0.0064 \\ 
\hspace{1.99cm}      $\mu_{I}=3.0$ Gev        & -0.0022 & -0.0031 & -0.0034 & -0.0038 & -0.0041 \\ 
$\mathcal{M}$ variation: $\mathcal{M}-20\,\%$ & 0.0013 & 0.0020 & 0.0022 & 0.0025 & 0.0028 \\ 
\hspace{2.02cm}          $\mathcal{M}+20\,\%$ & -0.0012 & -0.0018 & -0.0019 & -0.0022 & -0.0024 \\\hline 
Theoretical Error                             & 0.0071 & 0.0107 & 0.0124 & 0.0137 & 0.0151 \\\hline\hline 
$\alpha_0$                                    & 0.4267 & 0.4632 & 0.4789 & 0.4839 & 0.4857 \\\hline 
Experimental Error                            & 0.0082 & 0.0064 & 0.0067 & 0.0069 & 0.0070 \\\hline 
$x_{\mu}$ variation: $x_{\mu}=0.5$            & 0.0052 & 0.0027 & -0.0010 & -0.0035 & -0.0054 \\ 
\hspace{2cm}         $x_{\mu}=2.0$            & -0.0029 & -0.0021 & 0.0007 & 0.0027 & 0.0042 \\ 
$\mathcal{M}$ variation: $\mathcal{M}-20\,\%$ & 0.0324 & 0.0359 & 0.0377 & 0.0376 & 0.0366 \\ 
\hspace{2.02cm}          $\mathcal{M}+20\,\%$ & -0.0236 & -0.0266 & -0.0268 & -0.0283 & -0.0278 \\\hline 
Theoretical Error                             & 0.0328 & 0.0360 & 0.0377 & 0.0377 & 0.0370 \\\hline 
\end{tabular}
\caption{Results for $\alpha_{s}(Q)$ and $\alpha_{0}$ as obtained from fits to $C$ moments measured by JADE and OPAL for centre-of-mass energies between 14.0 and 206.6 GeV using theoretical NLO predictions.} 
\label{table:CfitresultNLO} 
\end{table}

\begin{table}[h!]
\centering
\begin{tabular}{{l|ccccc}}
NNLO &$\left<C\right>$ & $\left<C^2\right>$ & $\left<C^3\right>$ & $\left<C^4\right>$ & $\left<C^5\right>$\\\hline\hline 
$\chi^{2}/\mathrm{dof}$                       & 1.1574 & 1.2418 & 1.2353 & 1.1735 & 1.0216 \\\hline 
$\alpha_s(M_Z)$                               & 0.1161 & 0.1180 & 0.1193 & 0.1202 & 0.1216 \\\hline 
Experimental Error                            & 0.0011 & 0.0013 & 0.0016 & 0.0017 & 0.0019 \\\hline 
$x_{\mu}$ variation: $x_{\mu}=0.5$            & -0.0020 & -0.0033 & -0.0039 & -0.0043 & -0.0046 \\ 
\hspace{2cm}         $x_{\mu}=2.0$            & 0.0025 & 0.0040 & 0.0047 & 0.0051 & 0.0056 \\ 
$\mu_{I}$ variation: $\mu_{I}=1.0$ Gev        & 0.0019 & 0.0022 & 0.0023 & 0.0025 & 0.0028 \\ 
\hspace{1.99cm}      $\mu_{I}=3.0$ Gev        & -0.0013 & -0.0014 & -0.0015 & -0.0016 & -0.0017 \\ 
$\mathcal{M}$ variation: $\mathcal{M}-20\,\%$ & 0.0011 & 0.0012 & 0.0013 & 0.0014 & 0.0015 \\ 
\hspace{2.02cm}          $\mathcal{M}+20\,\%$ & -0.0010 & -0.0011 & -0.0012 & -0.0013 & -0.0014 \\\hline 
Theoretical Error                             & 0.0033 & 0.0047 & 0.0054 & 0.0059 & 0.0064 \\\hline\hline 
$\alpha_0$                                    & 0.4689 & 0.4897 & 0.4919 & 0.4877 & 0.4828 \\\hline 
Experimental Error                            & 0.0071 & 0.0063 & 0.0067 & 0.0069 & 0.0070 \\\hline 
$x_{\mu}$ variation: $x_{\mu}=0.5$            & 0.0166 & 0.0095 & 0.0053 & 0.0027 & 0.0010 \\ 
\hspace{2cm}         $x_{\mu}=2.0$            & -0.0105 & -0.0066 & -0.0033 & -0.0013 & 0.0001 \\ 
$\mathcal{M}$ variation: $\mathcal{M}-20\,\%$ & 0.0316 & 0.0360 & 0.0359 & 0.0346 & 0.0326 \\ 
\hspace{2.02cm}          $\mathcal{M}+20\,\%$ & -0.0234 & -0.0264 & -0.0265 & -0.0258 & -0.0246 \\\hline 
Theoretical Error                             & 0.0357 & 0.0372 & 0.0363 & 0.0347 & 0.0326 \\\hline 
\end{tabular}
\caption{Results for $\alpha_{s}(Q)$ and $\alpha_{0}$ as obtained from fits to $C$ moments measured by JADE and OPAL for centre-of-mass energies between 14.0 and 206.6 GeV using theoretical NNLO predictions.} 
\label{table:CfitresultNNLO} 
\end{table}

\begin{table}[h!]
\centering
\begin{tabular}{{l|ccccc}}
NLO &$\left<\rho\right>$ & $\left<\rho^2\right>$\\\hline\hline 
$\chi^{2}/\mathrm{dof}$                       & 0.6587 & 0.7547 \\\hline 
$\alpha_s(M_Z)$                               & 0.1164 & 0.1152 \\\hline 
Experimental Error                            & 0.0023 & 0.0033 \\\hline 
$x_{\mu}$ variation: $x_{\mu}=0.5$            & -0.0028 & -0.0038 \\ 
\hspace{2cm}         $x_{\mu}=2.0$            & 0.0039 & 0.0049 \\ 
$\mu_{I}$ variation: $\mu_{I}=1.0$ Gev        & 0.0014 & 0.0014 \\ 
\hspace{1.99cm}      $\mu_{I}=3.0$ Gev        & -0.0011 & -0.0011 \\ 
$\mathcal{M}$ variation: $\mathcal{M}-20\,\%$ & 0.0006 & 0.0006 \\ 
\hspace{2.02cm}          $\mathcal{M}+20\,\%$ & -0.0006 & -0.0006 \\\hline 
Theoretical Error                             & 0.0042 & 0.0051 \\\hline\hline 
$\alpha_0$                                    & 0.5914 & 0.5657 \\\hline 
Experimental Error                            & 0.0268 & 0.0361 \\\hline 
$x_{\mu}$ variation: $x_{\mu}=0.5$            & 0.0115 & 0.0092 \\ 
\hspace{2cm}         $x_{\mu}=2.0$            & -0.0042 & -0.0047 \\ 
$\mathcal{M}$ variation: $\mathcal{M}-20\,\%$ & 0.0795 & 0.0748 \\ 
\hspace{2.02cm}          $\mathcal{M}+20\,\%$ & -0.0539 & -0.0508 \\\hline 
Theoretical Error                             & 0.0803 & 0.0753 \\\hline 
\end{tabular}
\caption{Results for $\alpha_{s}(Q)$ and $\alpha_{0}$ as obtained from fits to $\rho$ moments measured by JADE and OPAL for centre-of-mass energies between 14.0 and 206.6 GeV using theoretical NLO predictions.} 
\label{table:MHfitresultNLO} 
\end{table}

\begin{table}[h!]
\centering
\begin{tabular}{{l|ccccc}}
NNLO &$\left<\rho\right>$ & $\left<\rho^2\right>$\\\hline\hline 
$\chi^{2}/\mathrm{dof}$                       & 0.6750 & 0.7607 \\\hline 
$\alpha_s(M_Z)$                               & 0.1142 & 0.1113 \\\hline 
Experimental Error                            & 0.0021 & 0.0030 \\\hline 
$x_{\mu}$ variation: $x_{\mu}=0.5$            & -0.0009 & -0.0012 \\ 
\hspace{2cm}         $x_{\mu}=2.0$            & 0.0013 & 0.0017 \\ 
$\mu_{I}$ variation: $\mu_{I}=1.0$ Gev        & 0.0012 & 0.0010 \\ 
\hspace{1.99cm}      $\mu_{I}=3.0$ Gev        & -0.0008 & -0.0007 \\ 
$\mathcal{M}$ variation: $\mathcal{M}-20\,\%$ & 0.0007 & 0.0006 \\ 
\hspace{2.02cm}          $\mathcal{M}+20\,\%$ & -0.0006 & -0.0006 \\\hline 
Theoretical Error                             & 0.0018 & 0.0020 \\\hline\hline 
$\alpha_0$                                    & 0.6565 & 0.6208 \\\hline 
Experimental Error                            & 0.0224 & 0.0316 \\\hline 
$x_{\mu}$ variation: $x_{\mu}=0.5$            & 0.0312 & 0.0233 \\ 
\hspace{2cm}         $x_{\mu}=2.0$            & -0.0184 & -0.0143 \\ 
$\mathcal{M}$ variation: $\mathcal{M}-20\,\%$ & 0.0799 & 0.0759 \\ 
\hspace{2.02cm}          $\mathcal{M}+20\,\%$ & -0.0547 & -0.0518 \\\hline 
Theoretical Error                             & 0.0858 & 0.0794 \\\hline 
\end{tabular}
\caption{Results for $\alpha_{s}(Q)$ and $\alpha_{0}$ as obtained from fits to $\rho$ moments measured by JADE and OPAL for centre-of-mass energies between 14.0 and 206.6 GeV using theoretical NNLO predictions.} 
\label{table:MHfitresultNNLO} 
\end{table}

\begin{table}[h!]
\centering
\begin{tabular}{{l|ccccc}}
NLO &$\left<Y_3\right>$ & $\left<Y_3^2\right>$ & $\left<Y_3^3\right>$ & $\left<Y_3^4\right>$ & $\left<Y_3^5\right>$\\\hline\hline 
$\chi^{2}/\mathrm{dof}$                       & 0.8616 & 0.6386 & 0.7771 & 0.8691 & 0.9499 \\\hline 
$\alpha_s(M_Z)$                               & 0.1183 & 0.1172 & 0.1165 & 0.1149 & 0.1124 \\\hline 
Experimental Error                            & 0.0011 & 0.0016 & 0.0020 & 0.0026 & 0.0033 \\\hline 
$x_{\mu}$ variation: $x_{\mu}=0.5$            & -0.0040 & -0.0042 & -0.0040 & -0.0038 & -0.0035 \\ 
\hspace{2cm}         $x_{\mu}=2.0$            & 0.0053 & 0.0054 & 0.0052 & 0.0049 & 0.0045 \\\hline 
Theoretical Error                             & 0.0047 & 0.0048 & 0.0046 & 0.0043 & 0.0040 \\\hline 
\end{tabular}
\caption{Results for $\alpha_{s}(Q)$ and $\alpha_{0}$ as obtained from fits to $Y_3$ moments measured by JADE and OPAL for centre-of-mass energies between 14.0 and 206.6 GeV using theoretical NLO predictions.} 
\label{table:Y3fitresultNLO} 
\end{table}

\begin{table}[h!]
\centering
\begin{tabular}{{l|ccccc}}
NNLO &$\left<Y_3\right>$ & $\left<Y_3^2\right>$ & $\left<Y_3^3\right>$ & $\left<Y_3^4\right>$ & $\left<Y_3^5\right>$\\\hline\hline 
$\chi^{2}/\mathrm{dof}$                       & 0.8577 & 0.6581 & 0.7948 & 0.8781 & 0.9557 \\\hline 
$\alpha_s(M_Z)$                               & 0.1156 & 0.1136 & 0.1136 & 0.1129 & 0.1106 \\\hline 
Experimental Error                            & 0.0010 & 0.0015 & 0.0019 & 0.0025 & 0.0032 \\\hline 
$x_{\mu}$ variation: $x_{\mu}=0.5$            & -0.0005 & -0.0008 & -0.0006 & -0.0002 & -0.0002 \\ 
\hspace{2cm}         $x_{\mu}=2.0$            & 0.0015 & 0.0017 & 0.0015 & 0.0013 & 0.0012 \\\hline 
Theoretical Error                             & 0.0010 & 0.0013 & 0.0011 & 0.0008 & 0.0007 \\\hline 
\end{tabular}
\caption{Results for $\alpha_{s}(Q)$ and $\alpha_{0}$ as obtained from fits to $Y_3$ moments measured by JADE and OPAL for centre-of-mass energies between 14.0 and 206.6 GeV using theoretical NNLO predictions.} 
\label{table:Y3fitresultNNLO} 
\end{table}

\begin{table}[h!]
\centering
\begin{tabular}{{l|ccccc}}
NLO &$\left<B_T\right>$ & $\left<B_T^2\right>$ & $\left<B_T^3\right>$ & $\left<B_T^4\right>$ & $\left<B_T^5\right>$\\\hline\hline 
$\chi^{2}/\mathrm{dof}$                       & 1.5775 & 1.6741 & 1.5926 & 1.4005 & 1.1996 \\\hline 
$\alpha_s(M_Z)$                               & 0.1199 & 0.1276 & 0.1308 & 0.1327 & 0.1347 \\\hline 
Experimental Error                            & 0.0012 & 0.0018 & 0.0023 & 0.0027 & 0.0031 \\\hline 
$x_{\mu}$ variation: $x_{\mu}=0.5$            & -0.0037 & -0.0078 & -0.0093 & -0.0101 & -0.0108 \\ 
\hspace{2cm}         $x_{\mu}=2.0$            & 0.0049 & 0.0094 & 0.0112 & 0.0123 & 0.0133 \\ 
$\mu_{I}$ variation: $\mu_{I}=1.0$ Gev        & 0.0021 & 0.0032 & 0.0036 & 0.0038 & 0.0041 \\ 
\hspace{1.99cm}      $\mu_{I}=3.0$ Gev        & -0.0016 & -0.0024 & -0.0026 & -0.0028 & -0.0030 \\ 
$\mathcal{M}$ variation: $\mathcal{M}-20\,\%$ & 0.0010 & 0.0015 & 0.0016 & 0.0017 & 0.0018 \\ 
\hspace{2.02cm}          $\mathcal{M}+20\,\%$ & -0.0009 & -0.0014 & -0.0015 & -0.0016 & -0.0017 \\\hline 
Theoretical Error                             & 0.0054 & 0.0101 & 0.0119 & 0.0130 & 0.0140 \\\hline\hline 
$\alpha_0$                                    & 0.4252 & 0.4897 & 0.5180 & 0.5193 & 0.5088 \\\hline 
Experimental Error                            & 0.0130 & 0.0105 & 0.0112 & 0.0129 & 0.0146 \\\hline 
$x_{\mu}$ variation: $x_{\mu}=0.5$            & 0.0154 & 0.0083 & 0.0005 & -0.0050 & -0.0093 \\ 
\hspace{2cm}         $x_{\mu}=2.0$            & -0.0106 & -0.0074 & -0.0015 & 0.0031 & 0.0070 \\ 
$\mathcal{M}$ variation: $\mathcal{M}-20\,\%$ & 0.0333 & 0.0443 & 0.0500 & 0.0493 & 0.0452 \\ 
\hspace{2.02cm}          $\mathcal{M}+20\,\%$ & -0.0238 & -0.0318 & -0.0358 & -0.0354 & -0.0329 \\\hline 
Theoretical Error                             & 0.0367 & 0.0451 & 0.0501 & 0.0496 & 0.0462 \\\hline 
\end{tabular}
\caption{Results for $\alpha_{s}(Q)$ and $\alpha_{0}$ as obtained from fits to $B_T$ moments measured by JADE and OPAL for centre-of-mass energies between 14.0 and 206.6 GeV using theoretical NLO predictions.} 
\label{table:BTfitresultNLO} 
\end{table}

\begin{table}[h!]
\centering
\begin{tabular}{{l|ccccc}}
NNLO &$\left<B_T\right>$ & $\left<B_T^2\right>$ & $\left<B_T^3\right>$ & $\left<B_T^4\right>$ & $\left<B_T^5\right>$\\\hline\hline 
$\chi^{2}/\mathrm{dof}$                       & 1.6191 & 1.4765 & 1.3723 & 1.2059 & 1.0363 \\\hline 
$\alpha_s(M_Z)$                               & 0.1164 & 0.1158 & 0.1158 & 0.1158 & 0.1162 \\\hline 
Experimental Error                            & 0.0011 & 0.0014 & 0.0017 & 0.0019 & 0.0022 \\\hline 
$x_{\mu}$ variation: $x_{\mu}=0.5$            & -0.0012 & -0.0027 & -0.0033 & -0.0035 & -0.0037 \\ 
\hspace{2cm}         $x_{\mu}=2.0$            & 0.0016 & 0.0035 & 0.0041 & 0.0044 & 0.0047 \\ 
$\mu_{I}$ variation: $\mu_{I}=1.0$ Gev        & 0.0017 & 0.0017 & 0.0018 & 0.0018 & 0.0019 \\ 
\hspace{1.99cm}      $\mu_{I}=3.0$ Gev        & -0.0012 & -0.0011 & -0.0012 & -0.0012 & -0.0012 \\ 
$\mathcal{M}$ variation: $\mathcal{M}-20\,\%$ & 0.0010 & 0.0010 & 0.0010 & 0.0010 & 0.0011 \\ 
\hspace{2.02cm}          $\mathcal{M}+20\,\%$ & -0.0009 & -0.0009 & -0.0009 & -0.0009 & -0.0010 \\\hline 
Theoretical Error                             & 0.0025 & 0.0040 & 0.0046 & 0.0049 & 0.0052 \\\hline\hline 
$\alpha_0$                                    & 0.4844 & 0.5053 & 0.5059 & 0.4938 & 0.4772 \\\hline 
Experimental Error                            & 0.0104 & 0.0094 & 0.0098 & 0.0108 & 0.0117 \\\hline 
$x_{\mu}$ variation: $x_{\mu}=0.5$            & 0.0491 & 0.0272 & 0.0190 & 0.0142 & 0.0109 \\ 
\hspace{2cm}         $x_{\mu}=2.0$            & -0.0295 & -0.0186 & -0.0129 & -0.0093 & -0.0066 \\ 
$\mathcal{M}$ variation: $\mathcal{M}-20\,\%$ & 0.0325 & 0.0419 & 0.0436 & 0.0415 & 0.0374 \\ 
\hspace{2.02cm}          $\mathcal{M}+20\,\%$ & -0.0240 & -0.0300 & -0.0312 & -0.0297 & -0.0270 \\\hline 
Theoretical Error                             & 0.0589 & 0.0500 & 0.0476 & 0.0438 & 0.0390 \\\hline 
\end{tabular}
\caption{Results for $\alpha_{s}(Q)$ and $\alpha_{0}$ as obtained from fits to $B_T$ moments measured by JADE and OPAL for centre-of-mass energies between 14.0 and 206.6 GeV using theoretical NNLO predictions.} 
\label{table:BTfitresultNNLO} 
\end{table}

\begin{table}[h!]
\centering
\begin{tabular}{{l|ccccc}}
NLO &$\left<B_W\right>$ & $\left<B_W^2\right>$ & $\left<B_W^3\right>$ & $\left<B_W^4\right>$ & $\left<B_W^5\right>$\\\hline\hline 
$\chi^{2}/\mathrm{dof}$                       & 1.5082 & 1.2870 & 1.1182 & 0.8965 & 0.6999 \\\hline 
$\alpha_s(M_Z)$                               & 0.1128 & 0.1077 & 0.1049 & 0.1023 & 0.1010 \\\hline 
Experimental Error                            & 0.0015 & 0.0020 & 0.0027 & 0.0033 & 0.0039 \\\hline 
$x_{\mu}$ variation: $x_{\mu}=0.5$            & 0.0007 & -0.0028 & -0.0026 & -0.0022 & -0.0019 \\ 
\hspace{2cm}         $x_{\mu}=2.0$            & 0.0006 & 0.0036 & 0.0035 & 0.0030 & 0.0027 \\ 
$\mu_{I}$ variation: $\mu_{I}=1.0$ Gev        & 0.0018 & 0.0015 & 0.0016 & 0.0016 & 0.0016 \\ 
\hspace{1.99cm}      $\mu_{I}=3.0$ Gev        & -0.0014 & -0.0012 & -0.0012 & -0.0012 & -0.0013 \\ 
$\mathcal{M}$ variation: $\mathcal{M}-20\,\%$ & 0.0008 & 0.0007 & 0.0007 & 0.0007 & 0.0008 \\ 
\hspace{2.02cm}          $\mathcal{M}+20\,\%$ & -0.0008 & -0.0007 & -0.0007 & -0.0007 & -0.0007 \\\hline 
Theoretical Error                             & 0.0021 & 0.0040 & 0.0039 & 0.0035 & 0.0032 \\\hline\hline 
$\alpha_0$                                    & 0.3960 & 0.3552 & 0.3090 & 0.2550 & 0.2025 \\\hline 
Experimental Error                            & 0.0106 & 0.0132 & 0.0154 & 0.0180 & 0.0203 \\\hline 
$x_{\mu}$ variation: $x_{\mu}=0.5$            & 0.0870 & 0.0256 & 0.0137 & 0.0107 & 0.0089 \\ 
\hspace{2cm}         $x_{\mu}=2.0$            & -0.0401 & -0.0166 & -0.0097 & -0.0074 & -0.0059 \\ 
$\mathcal{M}$ variation: $\mathcal{M}-20\,\%$ & 0.0357 & 0.0308 & 0.0222 & 0.0112 & -0.0005 \\ 
\hspace{2.02cm}          $\mathcal{M}+20\,\%$ & -0.0250 & -0.0214 & -0.0157 & -0.0084 & -0.0006 \\\hline 
Theoretical Error                             & 0.0941 & 0.0400 & 0.0261 & 0.0155 & 0.0089 \\\hline 
\end{tabular}
\caption{Results for $\alpha_{s}(Q)$ and $\alpha_{0}$ as obtained from fits to $B_W$ moments measured by JADE and OPAL for centre-of-mass energies between 14.0 and 206.6 GeV using theoretical NLO predictions.} 
\label{table:BWfitresultNLO} 
\end{table}

\begin{table}[h!]
\centering
\begin{tabular}{{l|ccccc}}
NNLO &$\left<B_W\right>$ & $\left<B_W^2\right>$ & $\left<B_W^3\right>$ & $\left<B_W^4\right>$ & $\left<B_W^5\right>$\\\hline\hline 
$\chi^{2}/\mathrm{dof}$                       & 1.5645 & 1.2884 & 1.1158 & 0.8943 & 0.7145 \\\hline 
$\alpha_s(M_Z)$                               & 0.1117 & 0.1058 & 0.1032 & 0.1014 & 0.1037 \\\hline 
Experimental Error                            & 0.0014 & 0.0018 & 0.0025 & 0.0031 & 0.0038 \\\hline 
$x_{\mu}$ variation: $x_{\mu}=0.5$            & -0.0010 & -0.0007 & -0.0006 & -0.0003 & 0.0005 \\ 
\hspace{2cm}         $x_{\mu}=2.0$            & 0.0007 & 0.0011 & 0.0010 & 0.0007 & 0.0001 \\ 
$\mu_{I}$ variation: $\mu_{I}=1.0$ Gev        & 0.0015 & 0.0012 & 0.0012 & 0.0012 & 0.0016 \\ 
\hspace{1.99cm}      $\mu_{I}=3.0$ Gev        & -0.0010 & -0.0008 & -0.0008 & -0.0008 & -0.0011 \\ 
$\mathcal{M}$ variation: $\mathcal{M}-20\,\%$ & 0.0009 & 0.0007 & 0.0007 & 0.0007 & 0.0009 \\ 
\hspace{2.02cm}          $\mathcal{M}+20\,\%$ & -0.0008 & -0.0006 & -0.0007 & -0.0007 & -0.0009 \\\hline 
Theoretical Error                             & 0.0020 & 0.0018 & 0.0017 & 0.0016 & 0.0020 \\\hline\hline 
$\alpha_0$                                    & 0.4632 & 0.4029 & 0.3519 & 0.2992 & 0.2744 \\\hline 
Experimental Error                            & 0.0083 & 0.0110 & 0.0127 & 0.0143 & 0.0137 \\\hline 
$x_{\mu}$ variation: $x_{\mu}=0.5$            & 0.0393 & 0.0313 & 0.0256 & 0.0221 & 0.0358 \\ 
\hspace{2cm}         $x_{\mu}=2.0$            & -0.0384 & -0.0232 & -0.0181 & -0.0155 & -0.0220 \\ 
$\mathcal{M}$ variation: $\mathcal{M}-20\,\%$ & 0.0378 & 0.0314 & 0.0227 & 0.0122 & 0.0045 \\ 
\hspace{2.02cm}          $\mathcal{M}+20\,\%$ & -0.0270 & -0.0222 & -0.0163 & -0.0093 & -0.0045 \\\hline 
Theoretical Error                             & 0.0545 & 0.0443 & 0.0342 & 0.0252 & 0.0360 \\\hline 
\end{tabular}
\caption{Results for $\alpha_{s}(Q)$ and $\alpha_{0}$ as obtained from fits to $B_W$ moments measured by JADE and OPAL for centre-of-mass energies between 14.0 and 206.6 GeV using theoretical NNLO predictions.} 
\label{table:BWfitresultNNLO} 
\end{table}

\end{appendix}

\end{document}